\documentstyle[11pt,epsf]{article}
\title{EXACT FEW-PARTICLE EIGENSTATES IN PARTIALLY REDUCED QED}

\author{Jurij W. Darewych\\ Department of Physics and Astronomy, York
University,\\ Toronto, Ontario, M3J~1P3, Canada\\ and Askold Duviryak\\
Department for Metal and Alloy Theory,\\ Institute for Condensed Matter
Physics of NAS of Ukraine,\\ Lviv, UA-79011, Ukraine}

\date{28.II.2002\qquad ExctQED2.tex}

\newcommand{\im}{{\,{\rm i}\,}}
\newcommand{\lab}[1]{\label{#1}} \newcommand{\re}[1]{(\ref{#1})}

\newcommand{\B}[1]{\mbox{\boldmath$#1$}}
\newfont{\ssb}{cmssbx10 scaled \magstephalf}
\newcommand{\s}[1]{\mbox{\ssb #1}}
\newcommand{\Bs}[1]{\mbox{\scriptsize\boldmath${#1}$}}

\begin{document} 
\def\ds{\displaystyle}
\def\nn{{\nonumber}}
\def\xp{x^\prime}
\def\tp{t^\prime}
\def\di{\partial}
\def\ni{\noindent}
\def\overphi{\overline \phi}
\def\overpsi{\overline \psi}
\def\overg{\overline g}
\def\overq{\overline q}
\def\overs{\overline s}
\def\overu{\overline u}
\def\overv{\overline v}
\def\overt{\overline t}
\def\overU{\overline U}
\def\overV{\overline V}
\def\dotphi{\dot\phi}
\def\dotchi{\dot\chi}
\def\bx{{\B x}}
\def\bxp{{{\bx}^\prime}}
\def\bk{{\B k}}
\def\br{{\B r}}
\def\brp{{\br}^{\prime}}
\def\bra{{\langle}}
\def\ket{{\rangle}}
\def\arrowal{\B\alpha}
\def\arrowsig{\B\sigma}
\def\aln{{\left ( {\alpha \over n} \right )}}
\def\be{ \begin{equation} }
\def\ee{ \end{equation} }

\maketitle


\begin{abstract}
We consider a reformulation of QED in which covariant Green  functions are
used to solve for the electromagnetic field in terms of the fermion fields.
It is shown that exact few-fermion eigenstates of the resulting
Hamiltonian can be obtained in the canonical equal-time formalism for the
case where there are no free photons. These eigenstates lead to two- and
three-body Dirac-like equations with electromagnetic interactions.
Perturbative and some numerical solutions of the two-body equations
are presented for positronium and muonium-like systems, for various
strengths of  the coupling.
\end{abstract}


\setcounter{equation}{0}
\renewcommand{\theequation}{1-\arabic{equation}}

\section{Introduction}

It has been pointed out in previous publications \cite{DarLB} that various
models in Quantum Field Theory (QFT), including QED, can be reformulated,
using mediating-field Green functions, in such a way that exact
few-particle eigenstates of the resulting partially truncated
Hamiltonian can be obtained.  This approach was then applied to
two-body eigenstates in the scalar Yukawa (Wick-Cutkosky) theory
\cite{Dar98-1,DB98}. We implement such an approach to QED in this paper.

The Lagrangian of two fermion fields, $\psi(x)$ and $\phi(x)$, interacting
electromagnetically, is
%
\begin{eqnarray}
{\cal L} &=&{\overline \psi}(x) \left( \im\gamma^\mu\,\di_\mu - q_1
\gamma^\mu A_\mu(x) - m_1\right) \psi (x) + {\overline \phi}(x) \left(
\im\gamma^\mu\,\di_\mu - q_2 \gamma^\mu  A_\mu(x) - m_2
\right) \phi (x) \nn \\  \label{1.1}
&&{}-{1\over 4}\; (\di_\alpha A_\beta (x)-\di_\beta A_\alpha (x)) \;
(\di^\alpha A^\beta(x)- \di^\beta A^\alpha (x) ).
\end{eqnarray}
The corresponding Euler-Lagrange equations of motion are the coupled
Dirac-Maxwell equations,
%
\begin{equation}\label{1.2}
(\im \gamma^\mu \partial_\mu - m_1) \psi(x) = q_1 \;\gamma^\mu A_\mu(x)
\psi (x),
\end{equation}
\begin{equation}\label{1.3}
(\im \gamma^\mu \partial_\mu - m_2) \phi(x) = q_2 \;\gamma^\mu A_\mu(x)
\phi (x),
\end{equation}
and
\begin{equation}\label{1.4}
\partial_\mu \di^\mu A^\nu (x) - \di^\nu \di_\mu A^\mu (x) =  j^\nu(x),
\end{equation}
where
\begin{equation}\label{1.5}
j^\nu(x) = q_1\; \overline \psi (x) \gamma^\nu \psi(x) + q_2\;
\overline \phi (x)\gamma^\nu \phi(x).
\end{equation}
The equations \re{1.2} - \re{1.4} can be decoupled in part by using the
well-known \cite{CED,Barut} formal solution of the Maxwell equation
\re{1.4}, namely
%
\begin{equation}\label{1.6}
A_\mu(x) = A_\mu^0 (x) +  \int  D_{\mu \nu} (x - x^\prime)\;  j^\nu (\xp)
d^4x^\prime,
\end{equation}
where  $D_{\mu \nu} (x-\xp)$ is a  Green function (or photon propagator in
QFT terminology), defined by
%
\begin{equation}\label{1.7}
\di_\alpha \di^\alpha D_{\mu \nu} (x-\xp) - \di_\mu \di^\alpha D_{\alpha
\nu} (x - x^\prime) = g_{\mu \nu} \delta^4(x-\xp) ,
\end{equation}
and $A_\mu^0(x)$ is a solution of the homogeneous (or ``free field'')
equation \re{1.4} with $j^\mu (x) = 0$.

We recall, in passing, that equation \re{1.7} does not define the covariant
Green function $D_{\mu \nu} (x-\xp)$ uniquely. For one thing, one can always
add a solution of the homogeneous equation ({\sl Eq.} \re{1.7} with $g_{\mu
\nu} \to 0)$. This allows for a certain freedom in the choice of $D_{\mu
\nu}$,  as is discussed in standard texts (e.g. refs. \cite{CED,Barut}). In
practice, the solution of Eq. \re{1.7}, like that of Eq. \re{1.4}, requires
a choice of gauge. However, we do not need to specify one at this stage.

Substitution of the formal solution \re{1.6} into equations \re{1.2} and
\re{1.3} yields the ``partially reduced" equations,
%
\begin{equation}\label{1.8}
( \im \gamma^\mu \di_\mu - m_1) \psi(x) = q_1 \gamma^\mu  \left( A_\mu^0(x)
 +  \int d^4\xp D_{\mu \nu}(x-\xp) j^\nu(\xp) \right) \psi(x) ,
\end{equation}
and
%
\begin{equation}\label{1.9}
( \im \gamma^\mu \di_\mu - m_2) \phi(x) = q_2 \gamma^\mu  \left( A_\mu^0(x)
 +  \int d^4\xp D_{\mu \nu} (x-\xp) j^\nu(\xp) \right) \phi(x) .
\end{equation}
These are nonlinear coupled Dirac equations for two different fermion
fields. To our knowledge no exact (analytic or numeric) solutions of
equations \re{1.8} and \re{1.9} for classical fields have been reported in
the literature, even for the  case of a single fermion field (say $\phi
=0$), though approximate  (perturbative) solutions have
been discussed by various authors, particularly Barut and his co-workers
(see refs. \cite{Grandy,Barut89} and citations therein). However, our
interest here is in the  quantized field theory.

The partially reduced equations \re{1.8}-\re{1.9}
are derivable from the stationary action principle
%
\begin{equation}\label{1.10}
\delta\;S[\psi,\phi] = \delta \int {\cal L}_{R}\; d^4x = 0,
\end{equation}
with the Lagrangian density
%
\begin{eqnarray}
{\cal L}_R & =&  \overpsi(x) \left( \im \gamma^\mu \di_\mu - m_1 - q_1
\gamma_\mu A^\mu_0(x) \right) \psi(x) + \overphi(x) ( \im \gamma^\mu
\di_\mu - m_2 - q_2 \gamma_\mu A^\mu_0(x)) \phi(x)  \nn \\
\label{1.11}
& & - {1\over 2}  \int d^4\xp j^\mu(\xp) D_{\mu \nu}(x-\xp) j^\nu(x),
\end{eqnarray}
provided that the Green function is symmetric in the sense that
%
\begin{equation}\label{1.12}
D_{\mu \nu}(x-\xp) = D_{\mu \nu}(\xp-x) \qquad {\rm and} \qquad
D_{\mu \nu}(x-\xp) = D_{\nu \mu}(x-\xp) .
\end{equation}

One can proceed to do conventional covariant perturbation
theory using the reformulated QED
Lagrangian \re{1.11}. The  interaction part of \re{1.11}
has a somewhat modified structure from that of the
 usual formulation of QED.  Thus, there are two interaction terms.
The last term of \re{1.11} is a
``current-current'' interaction which contains the  photon
propagator sandwiched between the fermionic
currents.  As such, it corresponds to Feynman diagrams
without external photon lines.
The terms containing $A^\mu_0$ corresponds to diagrams
that cannot be generated by the
 term containing  $D_{\mu \nu}$, particularly diagrams
involving external photon lines (care would have to
be taken not to double count physical effects).
 However, we shall not pursue perturbation theory in this work.
Rather, we shall consider an approach that allows one to write
down some unorthodox but exact eigenstates of a truncated model, in which
terms involving $A^\nu_0$ are ignored.

The paper is organized as follows. In section 2 we quantize the system
using the canonical equal time formalism in the  Schr\"odinger picture.
In section 3 an unconventional ``empty"  vacuum state
is used  to construct exact one-,
two-, and  three-fermion eigenstates of the Hamiltonian, truncated to
exclude states with free (physical) photons.
 In section 4 we show that the resulting two-fermion equation is the Breit
equation in the Coulomb gauge, but that it is the Eddington-Gaunt
equation in the Lorentz
gauge. In  section 5 we demonstrate that the Breit equation
can be obtained in the Lorentz  gauge, provided that
higher-order retardation effects are taken into account.

The reduction
of the Breit equation to radial form is described  briefly
in section 6. For states of
zero total angular momentum ($J=0$), four
coupled radial equations are shown to arise. The
analytical structure of their solutions is studied in section 7.
Perturbative $O(\alpha^4)$ corrections to the Rydberg spectrum
of $J=0$ states are obtained in section  8. In  the case of equal
rest masses the $J=0^+$ state equations have no unusual singularities and
can be solved numerically. Some of these results are  presented and
discussed in section 9. The remainder of the paper is devoted  to
the study of $J>0$ states. In the section 10 the set of eight
coupled  radial first-order differential equations is reduced
to four first-order  ones and then to two second-order
Schr\"odinger-like equations. They are  solved perturbatively in
section 11 and $O(\alpha^4)$
relativistic corrections to the non-relativistic mass
spectrum are  obtained. A summary and concluding
remarks are given in section 12.


\setcounter{equation}{0}
\renewcommand{\theequation}{2-\arabic{equation}}

\section{Hamiltonian in the canonical, equal-time formalism}

We consider this theory in the canonical, equal-time formalism. To this end
we write down the Hamiltonian density corresponding to the Lagrangian
\re{1.11}
%
\begin{eqnarray}
{\cal H}_R &=& \psi^{\dag}(x) ( -\im \arrowal \cdot \B\nabla + m_1\beta )
\psi(x) + q_1 \overpsi(x)\gamma_\mu A^\mu_0(x)) \psi(x) \nn \\ \label{2.1}
&&{} + \phi^{\dag}(x) ( -\im \arrowal \cdot \B\nabla + m_2\beta ) \phi(x) +
q_2 \overphi (x) \gamma_\mu A^\mu_0(x)) \phi(x) \\
&&{} + {1\over 2} \int d^4\xp\; j^\mu(\xp) D_{\mu \nu}(x-\xp) j^\nu(x) \nn
\end{eqnarray}
where we have not written out the Hamiltonian density for the free
$A_0^\mu(x)$ field.

Equal-time quantization corresponds to the imposition of
anticommutation rules for the fermion fields, namely
%
\begin{equation}\label{2.2}
\{\psi_\alpha({\bx},t),\psi^{\dag}_\beta({\B y},t)\} =
\{\phi_\alpha({\bx},t),\phi^{\dag}_\beta({\B y},t)\}
= \delta_{\alpha \beta} \delta^3({\B{x-y}}),
\end{equation}
and all others vanish. In addition, if $A^\mu_0 \ne 0$, there are the usual
commutation rules for the $A^\mu_0$ field, and commutation of the $A^\mu_0$
field operators with the $\psi$ and $\phi$ field operators.

The Hamiltonian \re{2.1} contains an interaction term that is nonlocal in
time, which can complicate the transition to a quantized theory.  We shall
avoid this problem by working in the  Schr\"odinger picture with $t=0$ in
the expressions for the field operators and currents, that is $\psi(x) =
\psi({\bx},t=0)$, $j^\mu(x) = j^\mu({\bx},t=0)$, etc. in equation \re{2.1}.
This corresponds to neglecting higher order retardation effects. Thereupon
we obtain the result
%
\begin{equation}\label{2.3}
\int d t^\prime \, D_{\mu \nu}(x-\xp) =  G_{\mu \nu} ({\bx - \bxp}),
\end{equation}
where
%
\begin{equation}\label{2.4}
G_{\mu \nu}(\bx) = \int {{d^3k} \over {(2\pi)^3}} G_{\mu \nu}(\bk)  e^{\im
\Bs k \cdot \Bs x}, \qquad {\rm and}\qquad G_{\mu \nu}(\bk) = D_{\mu
\nu}(k^\mu = (0,\bk)) .
\end{equation}
For example, in the Lorentz gauge ($\di_\mu A^{\mu} = 0$), we have
%
\begin{equation}\label{2.5}
G_{\mu \nu} (\bx) =  g_{\mu \nu} {1 \over { 4 \pi |\bx|}}.
\end{equation}
Thus, in the Schr\"odinger picture, the third term of the Hamiltonian
density \re{2.1} takes on the form
%
\begin{equation}\label{2.6}
{\cal H}_I({\bx}) = {1\over {2}} \int d^3x^\prime \; j^\mu({\bxp})
G_{\mu \nu} ({\bx-\bxp}) j^\nu ({\bx}).
\end{equation}

In the remainder of this paper, we shall consider the simplified model
without the interaction terms in \re{2.1} that contains $A_\mu^0$.
Such a model is suitable for describing few-fermion states interacting via
virtual photon exchange, but without decay or annihilation involving free
(physical) photons.  In short, in all that follows we consider the field
theory based on the Hamiltonian density of \re{2.1} but with $A^\mu_0(x)=0$.
An attractive feature of this model is that exact few-fermion eigenstates of
the Hamiltonian can be obtained.


\setcounter{equation}{0}
\renewcommand{\theequation}{3-\arabic{equation}}
\section{One, two and three fermion eigenstates}

We consider now the model for which the Hamiltonian, in the  Schr\"odinger
picture with $t=0$, is  given by the expression
%
\begin{equation}\label{3.1}
H_R = H_\psi + H_\phi + H_I,
\end{equation}
where
%
\begin{equation}\label{3.2}
H_\psi = \int d^3x \; \psi^{\dag}({\bx},0) (-\im \arrowal \cdot
\B\nabla +  m_1 \beta) \psi({\bx},0),  \end{equation}
%
%
\begin{equation}\label{3.3}
H_\phi = \int d^3x \; \phi^{\dag}({\bx},0) (-\im \arrowal \cdot \B\nabla
+  m_2 \beta) \phi({\bx},0),  \end{equation}
and $H_I=\int d^3x{\cal H}_I({\bx})$, where ${\cal H}_I({\bx})$ is given
in equation \re{2.6}. Note, again,  that the terms in $A^{\mu}_0$
have been suppressed, so that processes in which free (physical) photons
are emitted or absorbed are not accommodated.

The Hamiltonian $H_R$ has the same structure as the Coulomb-QED (CQED)
Hamiltonian, that is the Hamiltonian of QED in the Coulomb gauge, but with
the ``transverse-photon'' part (that contains $\arrowal \cdot {\B A}$)
turned off. Indeed $H_R$ would beidentical to $H_{CQED}$ if the indeces
$\mu$ and $\nu$ took on only the  value $0$ in equation \re{2.6} (as it is,
$\mu,\nu = 0,1,2,3$ in  Eq.\re{2.6}).   It has been shown earlier
\cite{DdiL96} that exact two-fermion eigenstates  of $H_{CQED}$ can be
written down if we use an unconventional (or  ``empty'') vacuum state,
$|\tilde 0 \ket$, defined by
%
\begin{equation}\label{3.4}
\psi_\alpha({\bx},0)|\tilde 0 \ket = \phi_\alpha({\bx},0)|\tilde 0 \ket =0 .
\end{equation}
The same is true of the present more realistic model, as we point out below.

The unconventional empty vacuum definition \re{3.4} means that
$\psi({\bx})$ is interpreted as a (free) Dirac-particle annihilation
operator, while   $\psi^{\dag}({\bx})$ is, correspondingly, a Dirac-particle
creation operator. By ``Dirac-particle'' we mean one described by the full
Dirac spinor, including positive and negative frequency components. (Recall
that in the conventional approach, {\sl i.e.} using a Dirac ``filled
negative energy sea'' vacuum which is annihilated by the positive frequency
component of $\psi$, it is only the negative-frequency component of $\psi$
that is an antiparticle creation operator, and the positive-frequency
component of $\psi^{\dag}$ that is the particle creation operator).

With these conventions, we write the normal-ordered Hamiltonian
%
\begin{eqnarray}
;H_R; = H_\psi + H_\phi + {1\over {2}} \int {d^3x \, d^3\xp} \,
 G_{\mu \nu} ({\bx - \bxp}) \,
\bigg[q_1^2 \overpsi \gamma^\mu (\overpsi^\prime \gamma^\nu \psi^\prime)
\psi  \nn \\ \label{3.5}
{} + q_1 q_2 \overphi \gamma^\mu (\overpsi^\prime \gamma^\nu \psi^\prime)
\phi
+ q_2 q_1 \overpsi \gamma^\mu (\overphi^\prime \gamma^\nu \phi^\prime) \psi
+ q_2^2 \overphi \gamma^\mu (\overphi^\prime \gamma^\nu \phi^\prime) \phi \bigg],
\end{eqnarray}
where $\psi = \psi({\bx})$ and $\overphi^\prime = \overphi  ({\bx}^\prime)$,
etc. The normal ordering is achieved by using the  anticommutation rules
\re{2.2} as usual; but note that it is not identical  to the conventional
normal ordering because of the unconventional empty  vacuum that is being
used, and the unconventional definition of $\psi$ and  $\phi$ as
annihilation operators and of $\psi^{\dag}, \phi^{\dag}$ as  creation
operators. To underscore this unconventional procedure we use the  notation
$;H_R;$ rather than $:H_R:$ in equation \re{3.5}.

We note that the state defined by
%
\begin{equation}\label{3.6}
|1\ket = \int d^3x \, \psi^{\dag}({\bx}) F({\bx}) |\tilde 0 \ket ,
\end{equation}
where $F({\bx})$ is a $4 \times 1$ c-number coefficient vector, is an
eigenstate of $;H_R;$ (Eq.  \re{3.5}) provided that  $F({\bx})$ satisfies
the equation
%
\begin{equation}\label{3.7}
(-\im \arrowal \cdot \B\nabla + m_1 \beta) F({\bx}) = E \, F({\bx}),
\end{equation}
which is the usual time-independent one-particle Dirac equation (with
positive and negative energy solutions), so that $F({\bx})$ is a Dirac
spinor. Therefore, we refer to $|1 \ket$ as   a one-Dirac-fermion state.

	Similarly, the two-Dirac-fermion state,
%
\begin{equation}\label{3.8}
|2 \ket = \int d^3x \, d^3y \, F_{\alpha\beta}({\bx},{\B y})
\psi_\alpha^{\dag}({\bx}) \phi_\beta^{\dag}({\B y}) |\tilde 0 \ket,
\end{equation}
(summation on $\alpha, \beta = 1,2,3,4$ implied) is an eigenstate of $;H_R;$
(equation  \re{3.5}) provided that the $4 \times 4$ eigenmatrix $F$
satisfies the equation
%
\begin{equation}\label{3.9}
h_{m_1}({\bx}) F({\bx},{\B y}) + \left [h_{m_2}({\B y}) F^T({\bx},{\B y})
\right ]^T + q_1 q_2 G_{\mu \nu}({\bx} - {\B y}) {\tilde \gamma}^\mu
F({\bx},{\B y}) ( {\tilde \gamma}^\nu)^T = E\, F({\bx},{\B y}),
\end{equation}
where $h_{m_j}({\bx}) = -\im \arrowal \cdot \B\nabla_{\Bs x} + m_j \beta$,
$\;\; {\tilde \gamma}^\mu = \gamma_0\, \gamma^\mu = (1, \arrowal)$, and the
superscript $T$  indicates the transpose of the matrix in question.

The detailed form of the interaction matrix $G_{\mu \nu} (\bx - \bx^\prime)$
depends on the choice of gauge. This equation \re{3.9} is a two-fermion
Dirac-like, or Breit-like, equation with positive and  negative energy
solutions, and is, in this respect, different from those obtained in the
conventional approach \cite{HarSuch85}-\cite{DD00}
in which the negative-energy solutions do not arise.

We note, in passing, that if the interaction is turned off in Eq. \re{3.9}
({\sl i.e.} $q_1=q_2=0$), then the solution can be written as
%
\begin{equation}\label{3.9a}
 F({\bx},{\B y}) = f({\bx}) g^T({\B y}) ,
\end{equation}
where $f({\bx})$ and $g({\B y})$ are solutions of the one-body Dirac
eigenvalue equation \re{3.7}. This indicates that in $F = [F_{ij}]$, the
index $i$ corresponds to  particle 1 (with coordinates ${\bx}$) while $j$
corresponds to particle 2  (with coordinates ${\B y}$).

In the rest frame of the two-fermion system ({\sl i.e.} when $|2 \ket$ is
taken to be an eigenstate of the momentum operator for this QFT, with
eigenvalue 0), equation \re{3.9} reduces to an analogous equation in the
single relative co-ordinate ${\br} = {\bx} -  {\B y}$:
%
\begin{equation}\label{3.10}
h_{m_1}({\br}) F({\br}) + \left [h_{m_2}(-{\br}) F^T({\br}) \right  ]^T +
q_1 q_2 G_{\mu \nu}({\br}) {\tilde \gamma}^\mu F({\br}) ({\tilde
\gamma}^\nu)^T = E \, F({\br}).
\end{equation}
It can, therefore, be reduced to a set of ordinary, coupled, first-order
differential equations for states of given $J^P$. Such equations can, at the
very least,  be solved numerically. This is a straightforward, though
somewhat tedious, problem \cite{DdiL96,Moor92} which we address below.

The structure of the Hamiltonian $;H_R;$ (Eq. \re{3.5}) is such that
generalizations to systems of more than two fermions are readily obtained.
Thus, the three fermion state, corresponding to a system like $|e^- e^-
\mu^+ \ket$, defined by
%
\begin{equation}\label{3.11}
|3 \ket = \int d^3x_1 d^3x_2 d^3x_3\; F_{\alpha_1 \alpha_2
\alpha_3}({\bx}_1,{\bx}_2,{\bx}_3) \;\psi^{\dag}_{\alpha_1}({\bx}_1)
\psi^{\dag}_{\alpha_2}({\bx}_2) \phi^{\dag}_{\alpha_3}({\bx}_3) |\tilde
0\ket,
\end{equation}
is an exact eigenstate of $;H_R;$ with eigenvalue $E$, provided that the
$4^3 = 64$ coefficient functions $F_{\alpha_1 \alpha_2
\alpha_3}({\bx}_1,{\bx}_2,{\bx}_3)$ satisfy the  three-body Dirac-like
equation,
%
\begin{eqnarray}
\left[ h_{m_1}({\bx}_1) \right]_{\alpha_1 \alpha}\; F_{\alpha \alpha_2
\alpha_3}({\bx}_1,{\bx}_2,{\bx}_3) + \left[h_{m_1}({\bx}_2)\right]_{\alpha_2
\alpha}\;F_{\alpha_1 \alpha \alpha_3}({\bx}_1,{\bx}_2,{\bx}_3)  \nn  \\
{}+ \left[h_{m_2}({\bx}_3) \right]_{\alpha_3 \alpha}\; F_{\alpha_1 \alpha_2
\alpha}({\bx}_1,{\bx}_2,{\bx}_3) + q_1 q_2 G_{\mu \nu} ({\bx}_3 - {\bx}_1)\;
({\tilde  \gamma}^\mu)_{\alpha_1 \alpha}\;({\tilde \gamma}^\nu)_{\alpha_3
\beta}\; F_{\alpha \alpha_2   \beta}({\bx}_1,{\bx}_2,{\bx}_3)     \nn  \\
\label{3.12}
{}+ q_1 q_2 G_{\mu \nu} ({\bx}_3 - {\bx}_2) \; ({\tilde
\gamma}^\mu)_{\alpha_2 \alpha} \; ({\tilde \gamma}^\nu)_{\alpha_3 \beta} \;
F_{\alpha_1 \alpha \beta}({\bx}_1,{\bx}_2,{\bx}_3)  \\
{}+  q_1^2 G_{\mu \nu} ({\bx}_1 - {\bx}_2) \; ({\tilde
\gamma}^\mu)_{\alpha_1 \alpha} \; ({\tilde \gamma}^\nu)_{\alpha_2 \beta} \;
F_{\alpha \beta \alpha_3  }({\bx}_1,{\bx}_2,{\bx}_3) \nn \\
{} = E \; F_{\alpha_1 \alpha_2 \alpha_3}({\bx}_1,{\bx}_2,{\bx}_3),   \nn
\end{eqnarray}
where summation on repeated spinor indices is implied.  In the rest frame of
the three-body system  equation \re{3.12} reduces to one in two independent
vectors only. Nevertheless,  the reduction of the equation for states of
given $J^P$ is more formidable than in the two-body case. Even then one is
left with the full complexity of a relativistic three-body system.  We shall
not consider solutions of the three-body equation \re{3.12} in this paper.


\setcounter{equation}{0}
\renewcommand{\theequation}{4-\arabic{equation}}

\section{Two-body equation in the Coulomb gauge}

At this stage we must specify the choice of Green's function, that is a choice of gauge.
We could use any gauge, in principle, but we shall use the Coulomb gauge, as this avoids
nonphysical degrees of freedom and the need to take account of  auxiliary conditions
(such as  $\di_\mu A^\mu = 0$ in the Lorentz gauge).  We use the relation
%
\begin{equation} \label{4.1}
D_{\mu \nu} (x) = \int {  {{d^4 k} \over {( 2\pi)^4}  } e^{- \im k \cdot x}
D_{\mu \nu} (k)},
\end{equation}
and note that
%
\begin{equation} \label{4.2}
D_{00} (k) =  {1 \over {\B k}^2}, \;\;\;\; D_{0 j} (k) = 0, \;\;\;\; D_{i
j} (k) =  {1 \over { k^2}} \left( \delta_{ij} - {{k_i k_j} \over {{\B  k}^2}
} \right)
\end{equation}
in the Coulomb gauge \cite{Sal52}. Therefore, if we use Eq. \re{2.3} and the
identity (\cite{BS57}, Eq. 39.8)
%
\begin{equation} \label{4.3}
\int {{d^3 k} \over {( 2\pi)^3}}\; e^{\im \Bs k \cdot \Bs x} \; {{{\B a}
\cdot \bk \;{\B b} \cdot \bk} \over {(\bk^2)^2}} = {1 \over {4\pi}} {1 \over
{2 r}} \left( {\B a} \cdot {\B b} - {{{\B a} \cdot \br \; {\B b} \cdot \br}
\over {r^2}} \right) ,
\end{equation}
we obtain the coordinate-space representation of the Coulomb-gauge Green
function
%
\begin{equation} \label{4.4}
G_{00}(\br) = {1 \over {4\pi r}}, \quad G_{0i} = 0 \quad {\rm} \quad
G_{ij}(\br) = - {1 \over {8\pi r}} \left( \delta_{ij} + {{x_i x_j} \over
r^2} \right) ,
\end{equation}
where $r=|\br|$ and $\br = (x_1, x_2, x_3)$.  Consequently, in the Coulomb
gauge, equation \re{3.10} becomes
%
\begin{equation} \label{4.5}
h_{m_1}(\br) F(\br) +\left[h_{m_2}(-\br) F^T(\br)\right]^T +  V(r) \left[
F(\br) - {1 \over 2} \left( \arrowal F(\br) \, \cdot \arrowal^T + {1 \over
r^2} \br \, \cdot \arrowal F(\br) \,  \br \, \cdot \arrowal^T \right)
\right] = E\, F(\br) ,
\end{equation}
where $\ds V(r) = {{q_1q_2} \over {4 \pi}}{1 \over r}$.

Keeping in mind the notation, as explained below Eq. \re{3.9a}, we see that
equation \re{4.5} is nothing other than the Breit equation \cite{Breit29},
written in the rest frame of the two-fermion system.

If we had used the Lorentz-gauge form of the Green function \re{2.5},
equation \re{3.10} would take on the form
%
\begin{equation} \label{4.6}
h_{m_1}(\br) F(\br) +\left[h_{m_2}(-\br) F^T(\br)\right]^T + V(r) \left[
F(\br) - \arrowal F(\br) \, \cdot \arrowal^T \right] = E\, F(\br) ,
\end{equation}
which is recognized to be the Eddington-Gaunt equation \cite{Ed29,Gaunt}.
The Gaunt equation, unlike the Breit equation, does not contain even
lowest-order retardation effects (see, also, section 5, following).
Therefore, it will yield energy eigenvalues that will differ from those of
the Breit equation already at $O(\alpha^4)$.

Note, also, that equations \re{4.5} and \re{4.6}, with the terms involving
$\arrowal$ left out, become identical to the Coulomb-QED model discussed
earlier \cite{DdiL96}.


\setcounter{equation}{0}
\renewcommand{\theequation}{5-\arabic{equation}}

\section{Two-body equation in the Lorentz gauge}

Although we shall use the Coulomb gauge in this paper, it is instructive to
see how, in the Lorentz gauge, one needs to keep retardation effects at
least to lowest non-vanishing order  in order to achieve the same results.
The Green function of the d'Alembert equation  in the Lorentz gauge has the
form:
%
\begin{equation}\lab{1}
D_{\mu\nu}(x) = g_{\mu\nu}D(x), \qquad  D(x) = \frac{1}{4\pi}\delta(x^2).
\end{equation}
The reduced Lagrangian ${\cal L}_R$ in this case reads (with $A_0(x)=0$):
%
\begin{eqnarray}\lab{2}
{\cal L}_R &=& {\cal L}_\psi + {\cal L}_\phi + {\cal L}_I
\nonumber\\
&=&\overpsi(x) ( \im \gamma^\mu \di_\mu - m_1) \psi(x)
+ \overphi(x) ( \im \gamma^\mu \di_\mu - m_2) \phi(x)
- {1\over 2} A^\mu(x) j_\mu(x),
\end{eqnarray}
where the potential of electromagnetic interparticle interaction is
%
\begin{equation}\lab{3}
A^\mu(x) = \int d^4\xp D(x-\xp) j^\mu(\xp) \equiv
\int d^3\xp \int d\tp D[(t-\tp)^2-(\bx-\bxp)^2] j^\mu(\tp,\bxp) .
\end{equation}
Thus,  the Lagrangian \re{2} is nonlocal in time. Because of this the
standard hamiltonization procedure is not applicable.

In order to employ the canonical Hamiltonian formalism it is necessary to
convert this Lagrangian to  single-time form. We shall do so by employing a
procedure which takes  into account the retardation effects approximately.

Using the substitution $\tp = t + \lambda$  in \re{3} and expanding the
current $j$ in a Taylor series in $\lambda$, we obtain the result
%
\begin{equation}\lab{4}
j(\bxp,\tp) = j(\bxp,t+\lambda) =
j(\bxp,t) +\lambda\dot j(\bxp,t) + \frac12\lambda^2\ddot j(\bxp,t) +
\cdots ,
\end{equation}
which reduces the potential \re{3} to the  form:
%
\begin{equation}\lab{5}
A^\mu(\bx,t) = \int d^3\xp \left\{G(r) j^\mu(\bxp,t)  + \frac12  G_1(r)\ddot
j^\mu(\bxp,t) + \cdots \right\},
\end{equation}
where $r = |\br| = |\bx-\bxp|$. The functions
%
\begin{equation}\lab{6}
G(r) = \int d\lambda \,  D(\lambda^2-r^2) = { 1 \over {4 \pi r}}
\end{equation}
and
%
\begin{equation}\lab{7}
G_1(r) = \int d\lambda \, D(\lambda^2-r^2)\lambda^2 = { r \over {4 \pi }}
\end{equation}
satisfy the relation
%
\begin{equation}\lab{8}
G^{\, \prime}_1(r) = rG(r).
\end{equation}
The terms of odd power in $\lambda$  in the expansion  \re{4} vanish,
because $D$ is an even function of $\lambda$ ({\sl c.f.} \re{6}, \re{7}).
As a result,  the interaction Lagrangian ${\cal L}_I$ (up to surface terms)
takes  the following single-time form:
%
\begin{equation}\lab{9}
{\cal L}_I \approx {\cal L}_I^{(0)} + {\cal L}_I^{(1)} + {\cal L}_I^{(2)},
\end{equation}
where
%
\begin{eqnarray}
{\cal L}_I^{(0)} &=& -\frac12\int d^3\xp G(r) j^0(\bx) j^0(\bxp),
\lab{10}\\
{\cal L}_I^{(1)} &=& \frac12\int d^3\xp G(r) \B j(\bx)\cdot\B j(\bxp),
\lab{11}\\
{\cal L}_I^{(2)} &=& \frac14\int d^3\xp G_1(r) \dot j^0(\bx) \dot j^0(\bxp)
\lab{12}
\end{eqnarray}
(hereafter we omit the common time argument $t$).

The quantized theory based on the Lagrangian ${\cal L}^{(0)} =
{\cal L}_\psi + {\cal L}_\phi + {\cal L}_I^{(0)}$ (the Coulomb QED
mentioned in section 3) was discussed in \cite{DdiL96}. It takes  into
account the relativistic kinematics of the fermion fields exactly, but
describes their electromagnetic interaction with the ``transverse-photon"
part turned off. The terms ${\cal L}_I^{(1)}$ and ${\cal L}_I^{(2)}$ can be
treated as first-order corrections to ${\cal L}^{(0)}$, thus providing the
approximate single-time form ${\cal L}_S = {\cal L}^{(0)} + {\cal L}_I^{(1)}
+ {\cal L}_I^{(2)}$ for the reduced nonlocal Lagrangian ${\cal L}_R$ ({\sl
c.f.} Eq. \re {1.11}). Other terms following from the expansion  are
corrections of higher order. They will not be considered in the present
paper.

The Lagrangian ${\cal L}_S$ leads to Euler-Lagrange equations which are
second-order in time derivatives, because of the term ${\cal L}_I^{(2)}$.
Thus, it describes the system with twice as many degrees of freedom
as ${\cal L}^{(0)}$ does, because $\psi^{\dag}$ are no longer the conjugate
momenta of $\psi$. This changes completely the dynamical content of the
fields $\psi$ and $\phi$. Since the second-order time derivatives occur in
small terms only,  they should be eliminated by means of the Euler-Lagrange
equations of a lower-order approximation. But the resulting field equations
are then not necessarily the Euler-Lagrange equations of a known Lagrangian.
Thus the transition to the Hamiltonian and to a canonical quantum
description becomes unclear.

To avoid this difficulty it is tempting to eliminate the time derivative of
the charge density $\dot j^0$ directly in ${\cal L}_I^{(2)}$ by taking into
account  the conservation law
%
\begin{equation}\lab{13}
\partial_\mu j^\mu = 0, \qquad  {\rm i.e.} \qquad \dot j^0 +
\B\nabla\cdot\B j = 0.
\end{equation}
This conservation law is a consequence of the Euler-Lagrange equations,
which follow  from the reduced Lagrangian ${\cal L}_R$ as well as from the
truncated one ${\cal L}^{(0)}$. However,  the direct use of the equations of
motion (or their consequences) in the Lagrangian is not a correct procedure:
it changes the equations of motion themselves. This fact was first
emphasized  in the case of the Golubenkov-Smorodinskii Lagrangian
\cite{18,19} and then subsequently discussed in the literature
\cite{20}-\cite{24}. Instead, one can use the method of ``double zero", used
in refs. \cite{20,23}.  In our case this consists of the following
modification of  the Lagrangian:
%
\begin{equation}\lab{14}
{\cal L}_S \rightarrow  {\overline{{\cal L}}_S} = {\cal L}_S + {\cal
L}_I^{(3)},
\end{equation}
where
%
\begin{equation}\lab{15}
{\cal L}_I^{(3)} = -\frac14\int d^3\xp G_1(r) \left\{\dot
j^0(\bx)+\B\nabla\cdot\B j(\bx)\right\}\left\{\dot
j^0(\bxp)+\B\nabla'\cdot\B j(\bxp)\right\}.
\end{equation}
It is easy to see that the term ${\cal L}_I^{(3)}$ possesses the property:
%
\begin{equation}\lab{16}
\delta\left.\int d^3x{\cal L}_I^{(3)}\right|_{\delta\int d^3x{\cal
L}^{(0)}=0}=0,
\end{equation}
so that it does not change the variational problem  to the accuracy required.
On the other hand, it cancels those terms of ${\cal L}_S$ which are
quadratic in time derivatives of the fields. Thus the modified Lagrangian
$\overline{{\cal L}}_S$ yields equations of motion, which are first order
in the time derivatives of the particle fields $\psi, \phi$.

Next, we perform the following transformation of the field variables:
%
\begin{eqnarray}
\psi \rightarrow \underline{\psi} = (1-\im q_1 W)\psi \approx e^{-\im q_1
W}\psi, &\qquad& \overpsi \rightarrow \underline{\overpsi} = (1+\im q_1
W)\overpsi \approx  e^{\im q_1 W}\overpsi,
\nonumber\\
\phi \rightarrow \underline{\phi} = (1-\im q_2 W)\phi  \approx e^{-\im q_2
W}\phi, &\qquad& \overphi \rightarrow \underline{\overphi} = (1+\im q_2
W)\overphi \approx  e^{\im q_2 W}\overphi,
\lab{17}
\end{eqnarray}
where
%
\begin{equation}\lab{18}
W(\bx) = \frac12\int d^3\xp G_1(r)\B\nabla'\cdot\B j(\bxp).
\end{equation}
The transformation \re{17} can be regarded as an approximate $U(1)$
gauge transformation, which however is not canonical
due to the dependence of $W$ on the fields. This transformation
removes time derivatives from the interaction part of $\overline{{\cal
L}}_S$. To the accuracy required, and up to surface terms,  Lagrangian
$\overline{{\cal L}}_S$ can be written in the form
%
\begin{equation}\lab{19}
\overline{{\cal L}}_S = \underline{\cal L}^{(0)} + \underline{\cal
L}_I^{(1)} + \frac14\int d^3\xp G_1(r)(\B\nabla\cdot\B{\underline{ j}}(\bx))
(\B\nabla'\cdot\B{\underline{j}}(\bxp)),
\end{equation}
where the notations $\underline{\cal L}^{(0)}$,  $\underline{ j}$, {\sl
etc.} mean that the fields $\psi$, $\phi$ are replaced by $\underline\psi$,
$\underline\phi$.

Integrating the last term of $\overline{{\cal L}}_S$ by parts, omitting
surface terms, and using equations \re{6} and  \re{8}, we reduce the
Lagrangian $\overline{{\cal L}}_S$ to the form
%
\begin{equation}\lab{20}
\overline{{\cal L}}_S = \im(\underline\psi^{\dag}\underline{\dot\psi} +
\underline\phi^{\dag}\underline{\dot\phi}) - \underline{\cal H}_S,
\end{equation}
where
%
\begin{eqnarray}\lab{21}
\underline{\cal H}_S &=&
\underline\psi^{\dag}(\bx) ( -\im \B\alpha\cdot \B\nabla + m_1\beta
)\underline\psi(\bx) + \underline\phi^{\dag}(\bx) ( -\im \B\alpha\cdot
\B\nabla + m_2\beta  )\underline\phi(\bx) \nonumber\\
&&\hspace{-3em}{}+\frac{1}{8\pi}\int d^3\xp \frac{1}{r}\left\{\underline
j^0(\bx)\underline j^0(\bxp) -\frac12\B{\underline{ j}}(\bx)
\cdot\B{\underline{ j}}(\bxp)-\frac{1}{2r^2}(\br\cdot\B{\underline{
j}}(\bx))(\br \cdot\B{\underline{ j}}(\bxp))\right\}.
\end{eqnarray}

This formulation allows us to treat the variables  $\underline\psi^{\dag}$
and $\underline\phi^{\dag}$as the canonical conjugates of $\underline\psi$
and $\underline\phi$, respectively. That is, we impose the anticommutation
relations \re{2.2} on the underscored fields, and not on the original ones,
when performing the quantization. Thereafter, since the Hamiltonian
$\underline H_S = \int d^3x \, \underline{\cal H}_S $ is formally identical
with the Coulomb gauge Hamiltonian ({\sl c.f.} \re{2.1} with \re{4.4}),
calculations like those of section 4 lead to the Breit equation \re{4.5}.


\setcounter{equation}{0}
\renewcommand{\theequation}{6-\arabic{equation}}

\section{Two-body equations in  block component form}

Although two fermion equations have been around since the 1920s, their full
reduction to radial form is of more recent vintage (see, for example
\cite{Malenf}, \cite{Grandy}). The reduction of  equation \re{4.5} to radial
form is essentially the same as presented in ref. \cite{DdiL96}, hence all
the details shall not be repeated here.   As shown in \cite{DdiL96}, we
note that Eq.\re{4.5} has the Schr\"odinger equation as a non-relativistic
limit, and  the Dirac equation   as a one-body limit if one of
$(m_1,m_2) \to \infty$.

For the two-fermion state $|2\ket$, Eq. \re{3.8}, to be simultaneously an
eigenstate of ${\B J}^2$, $J_3$, and parity, the ``bispinor" $F= [F_{ij}]$
must be of  the form
%
\begin{equation} \label{6-1}
F({\br})= {1\over r}\left[\matrix{\im s_1(r)\,\varphi^A({\hat
\br})+\im s_2(r)\,\varphi^0({\hat\br}),&t_1(r)\,\varphi^-({\hat
\br})+t_2(r)\,\varphi^+({\hat\br})\cr u_1(r)\,\varphi^-({\hat
\br})+u_2(r)\,\varphi^+({\hat\br}),&\im v_1(r)\,\varphi^A({\hat
\br})+\im v_2(r)\,\varphi^0({\hat\br}) \cr}\right],
\end{equation}
for $-(-1)^J=(-1)^{J\pm1}$ parity eigenstates, and
%
\begin{equation} \label{6-2}
F({\br})= {1\over r}\left[\matrix{\im s_1(r)\,\varphi^-({\hat
\br})+\im s_2(r)\,\varphi^+({\hat\br}),&t_1(r)\,\varphi^A({\hat
\br})+t_2(r)\,\varphi^0({\hat\br})\cr u_1(r)\,\varphi^A({\hat
\br})+u_2(r)\,\varphi^0({\hat\br}),&\im v_1(r)\,\varphi^-({\hat
\br})+\im v_2(r)\,\varphi^+({\hat\br}) \cr}\right],
\end{equation}
for $-(-1)^{J \pm 1}= (-1)^J$ parity eigenstates.

The $2 \times 2$ angular ``bispinor harmonics'' $\varphi^A({\hat\br})$,
$\varphi^0({\hat\br})$, $\varphi^+({\hat\br})$, $\varphi^-({\hat\br})$, for
given total angular  momentum quantum numbers $J$ and $m_J \equiv M$, are
%
\begin{equation} \label{6-3}
\varphi^A({\hat\br})= {1\over {\sqrt {2}}}Y^M_J({\hat\br})
\left[  \matrix{0&-1\cr 1&0}  \right],
\end{equation}
\begin{equation} \label{6-4}
\varphi^0({\hat\br}) = {1\over {\sqrt {2J(J+1)}}}
\left[  \matrix{{\sqrt{(J-M+1)(J+M)}}\,Y^{M-1}_J,&-M\,Y^M_J\cr
                -M\,Y^M_J,&-{\sqrt{(J+M+1)(J-
M)}}\,Y^{M+1}_J}  \right],
\end{equation}
\begin{equation} \label{6-5}
\varphi^+({\hat\br}) = {1\over {\sqrt {2J(2J-1)}}}
\left[\matrix{{\sqrt{(J+M-1)(J+M)}}\,Y^{M-1}_{J-
1},&{\sqrt{(J+M)(J-M)}}\,Y^M_{J-1}\cr
               {\sqrt{(J+M)(J-M)}}\,Y^M_{J-1},&{\sqrt{(J-M-
1)(J-M)}}\,Y^{M+1}_{J-1}}\right],
\end{equation}
and
\begin{eqnarray}
\varphi^-({\hat\br})  & = & {1\over {\sqrt {2(J+1)(2J+3)}}}   \label{6-6} \\
 & \times&  \left[\matrix{{\sqrt{(J-M+1)(J-M+2)}}Y^{M-1}_{J+
1},&-{\sqrt{(J+M+1)(J-M+1)}}Y^M_{J+1}\cr
               -{\sqrt{(J+M+1)(J-M+1)}}Y^M_{J+1},&
{\sqrt{(J+M+1)(J+M+2)}}Y^{M+1}_{J+1}}\right].  \nonumber
\end{eqnarray}
We note that $\varphi^A$ is antisymmetric and $\varphi^{0,\pm}$ are
symmetric matrices. Furthermore $\varphi^{0}$, $\varphi^{A}$ and
$\varphi^{\pm}$ correspond to opposite parity  because $Y^M_L(-  {\hat\br})
=  (-1)^L Y^M_L({\hat\br})$ and $\varphi^0$, $\varphi^A$ have $L= J$ whereas
$\varphi^{\pm}$ have $L= J \pm 1$. These four bispinor harmonics form an
orthonormal set, in the sense that $\int d\hat r \; {\rm Tr}(\varphi_{i}^{\,
\dag} \, \varphi_{j})=  \delta_{i\,j}$, where $i,j= A,0,+,-$ and the
integrations are taken over the entire solid angle.

The eight radial functions in the bispinors \re{6-1} and \re{6-2} are
solutions of the coupled radial equations that are obtained by substituting
\re{6-1},  \re{6-2} into equation \re{4.5} and equating the coefficients of
the four independent bispinor harmonics.

We make use of the following identities in carrying out the radial reduction:
%
\begin{equation} \label{6-7}
\arrowsig\B{ \cdot  p} f(r) \varphi({\hat{\br}}) =  - \im {df\over dr}\;
\arrowsig\cdot {\hat{\br}}\; \varphi({\hat{\br}}) + {\im\over r}
f(r)\arrowsig\cdot {\hat{\br}} \arrowsig\cdot {\B l} \;
\varphi({\hat{\br}}),
\end{equation}
where $\varphi({\hat{\br}})$ is any $2 \times 2$ bispinor harmonic,
$f(r)$ a radial function, and $\ds {{\hat{\br}} =  {{\br} \over r}}$,
and ${\B l} =  {\br} \times {\B p} =  - \im {\br} \times \B\nabla$.
In addition, we note the following useful properties of the above
bispinors harmonics:
%
\begin{eqnarray}
\arrowsig \cdot{\hat{\br}}\,\varphi^A &=& A\,\varphi^{-} -B\,\varphi^+,
\label{6-8} \\
\arrowsig\cdot{\hat{\br}}\,\varphi^0 &=&  B\,\varphi^{-} +A\,\varphi^+,
\label{6-9} \\
\arrowsig \cdot {\B l}\,\varphi^A &=& C\,\varphi^{0} , \label{6-10} \\
\arrowsig\cdot{\B l}\,\varphi^0 &=& -\varphi^{0}+C\,\varphi^A,
\label{6-11} \\
\arrowsig\cdot {\B l} \,\varphi^- &=& -(J+2)\varphi^{-}, \label{6-12} \\
\arrowsig\cdot  {\B l} \,\varphi^+ &= & (J-1)\varphi^{+}, \label{6-13}
\end{eqnarray}
where
%
\begin{equation} \label{6.14}
A = {\sqrt{{J+1\over 2\,J+1}}} \qquad
B= {\sqrt{{J\over 2\,J+1}}} \qquad
{\rm and} \qquad C= {\sqrt{J(J+1)}}.
\end{equation}

It is evident from equations \re{6-1} and \re{6-2} that, in general,
eight coupled radial equations are obtained, for arbitrary $J>0$.


\section{Radial reduction of the two-body equations for $J=0$ states}

\setcounter{equation}{0}
\renewcommand{\theequation}{7-\arabic{equation}}

For the $J= 0$ states, namely the $0^- (^1S_0)$ and $0^+ (^3P_0)$
states, only two linearly independent bispinor harmonics arise, namely
$\varphi^A$ and $\varphi^-$ (equations \re{6-3} and \re{6-6}), and so  $s_2=
t_2= u_2= v_2= 0$ in equations \re{6-1} and \re{6-2}. (Here, as  elsewhere,
we give in brackets the nonrelativistic limit designation,  $^{2S+1}L_J$,
corresponding to the $J^P$ state in question.) Thus there  is only one set
of four coupled radial equations for each of $0^-(^1S_0)$  and $0^+(^3P_0)$
states:
%
\begin{equation} \label{7-1}
(m_+ +V(r)-E)s(r) - t^{\prime}(r) - {K\over r} t(r) - u^{\prime}(r)  -
{K\over r} u(r) + \xi V(r) v(r) =0
\end{equation}
\begin{equation}  \label{7-2}
(m_- +V(r)-E)t(r) + s^{\prime}(r) - {K\over r} s(r) + v^{\prime}(r) -
{K\over r} v(r) + \eta V(r) u(r) =0
\end{equation}
\begin{equation}  \label{7-3}
(-m_-+V(r)-E)u(r) + s^{\prime}(r) - {K\over r} s(r) + v^{\prime}(r) -
{K\over r} v(r) + \eta V(r) t(r) =0
\end{equation}
\begin{equation}  \label{7-4}
(-m_+ +V(r)-E)v(r) - t^{\prime}(r) - {K\over r} t(r) - u^{\prime}(r) -
{K\over r} u(r) + \xi V(r) s(r) =0
\end{equation}
where $m_\pm = m_1 \pm m_2$,  $\ds {s^{\prime} = {{ds}\over{dr}}}$, the
potential $V(r) = - \alpha / r$ ($\alpha =|q_1 q_2|/4 \pi$), and
$ E$ is the eigenenergy (two-particle bound-state mass) to be determined,
while $K=1$ ($\xi=2, \eta=0$) for the $0^- (^1S_0)$ states and $K=-1$
($\xi=0, \eta=2$) for the $0^+ (^3P_0)$ states.  As shown in Ref.
\cite{DdiL96}, Eqs. \re{7-1} - \re{7-4} have the expected Schr\"odinger
nonrelativistic limit, and the Dirac one-body limit.

We note that the case with $\xi=\eta=0$ corresponds to the simplified  model
without transverse-photon interactions, that is the `Coulomb QED'  model of
Ref. \cite{DdiL96}. Similarly for $\xi=\eta=0$, if  the potential  is  $\ds
{V(r) = - {{q_1 q_2}\over {4 \pi}} {e^{-\mu r}\over r}}$, and the  sign of
the potential is reversed in \re{7-2} and \re{7-3} we recover the  $0^{\pm}$
radial equation of the Yukawa model  discussed in Ref.  \cite{Dar98-2}, for
which the inter fermion interaction is via a (massive  or massless) scalar
mediating field.

We should point out that equations \re{7-1} - \re{7-4}, like the Dirac
equations, have both positive and negative energy solutions. Indeed, in this
two-body case, there are solutions of four types: $E \simeq m_1 + m_2$,  $E
\simeq  -m_1 + m_2$, $E \simeq m_1 - m_2$ and $E \simeq -m_1 - m_2$,
as can be seen most easily from the $\alpha = 0$ case. Of these, two are
positive energy and two are negative energy solutions.

Since we do not have analytic solutions for the eigenenergies of the present
QED case, it is useful to illustrate this phenomenon on the scalar Yukawa
(or ``Wick-Cutkosky") model, in which scalar particles interact via a
massive or massless mediating field. For such a scalar model, analytic
expressions for the two-body bound state energy eigenvalues are available in
the massless-exchange case \cite{DB98}:
%
\begin{equation} \label{7-4a}
E =  \sqrt{m_1^2 + m_2^2 \pm 2 m_1 m_2 \sqrt {1-\aln^2} } ,
\end{equation}
where $\alpha$ is the effective dimensionless coupling constant, analogous
to the fine structure constant of QED, and $n$ is the principal quantum
number. The $\pm$ in Eq. \re{7-4a} correspond to  two segments of a
distorted semicircle. The upper branch of this distorted semicircle
corresponds to the upper (positive) sign in \re{7-4a}. It begins from
$E=m_1+m_2$ at $\alpha = 0$ (indeed,  $ E =  m_1 + m_2 - {1\over 2} m_r
{\aln}^2 - {1\over 8}m_r \left(1 + {m_r \over {m_+}}\right) \aln^4 + \cdots
\;\;, $ for ${\alpha \over n} \ll 1$), and decreases to a  $E =
\sqrt{m_1^2+m_2^2}$ at the critical value of $\alpha = n$, beyond which $E$
ceases to be real, and the wave functions cease to be normalizable.  The
lower branch, by contrast, begins from $E=|m_1-m_2|$ at $\alpha=0$ and rises
monotonically to meet the upper branch at the same critical point  $E \aln
= \sqrt{m_1^2+m_2^2}$ . These  $|m_1-m_2|$ type bound state eigenenergies
do not have the correct Balmer limit, since for this branch
%
\begin{equation} \label{7-4b}
E =  |m_1 - m_2| + {1\over 2} \left( {{m_1m_2}\over{|m_1-m_2|}}\right)
 {\aln}^2 + {1\over 8}  \left( {{m_1 m_2}\over{|m_1-m_2|}}\right)
\left(1 -  {{m_1 m_2} \over {(m_1-m_2)^2}}\right) \aln^4 + \cdots ,
\end{equation}
for $m_1 \ne m_2$, but $E = m \left( \aln + {1 \over 8} \aln^3 +
\cdots \right)$ for $m_1=m_2=m$.
Thus, this ``mixed energy" $E \simeq |m_1-m_2|$ bound-state spectrum must be
regarded as unphysical. There are also negative-energy solutions of the $E
\simeq -m_1 - m_2$ and $E \simeq -|m_1-m_2|$ type, but they are not
bound-states, since the potential effectively reverses sign for the
negative-energy solutions (as happens also in the Dirac-Coulomb case).
The same type of behaviour of the energy spectrum is observed in another
analytically solvable case, namely a fermion and a scalar particle
interacting via massless scalar quantum exchange \cite{ShD01}  Thus, we
expect that the energy eigenvalue spectrum of \re{7-1} - \re{7-4} will be
qualitatively similar to that of the scalar exchange models just described.

We have not been able to determine solutions to the coupled
radial equations \re{7-1}-\re{7-4} in terms of common analytic functions.
It is of interest, therefore, to consider the properties and
general behaviour of the solutions before commencing with numerical solutions.

In analogy with the scalar model just described, and with the Coulomb QED
case \cite{DdiL96} we expect that, as $\alpha$ increases, the eigenenergy
spectrum $E(\alpha)$ of equations \re{7-1}-\re{7-4} will have a qualitative
behaviour similar to that of the Dirac spectrum, namely that $E(\alpha)$
decreases monotonically from $E(\alpha= 0)= m_1+m_2$ until $\alpha$ hits a
critical value $\alpha_c$, beyond which $E(\alpha)$ ceases to be real. It is
possible to infer the value of $\alpha_c$  by considering the
ultra-relativistic limit, $p\to\infty$, in which case we can neglect the
masses $m_1$ and $m_2$, and seek solutions of \re{7-1}-\re{7-4} with $E=
m_1= m_2= 0$. (This approach, when applied to the one-body Dirac-Coulomb
case, yields the correct  critical values $\alpha_c = |\kappa| = |j+1/2|$.)
In this ultra-relativistic approximation, equations \re{7-1}-\re{7-4}  have
the solutions $t=u$, $s= v$, $|t|= |s|= 1$ (i.e. $\ds F\propto {1\over r}$)
with $\alpha^2_c = 4 K^2 / (1+\xi)(1+\eta)$, which gives $\alpha_c =
2/\sqrt{3}= 1.1547...$ for all $0^{\pm}$ states. Note however, that this
result does not mean that the value of the two-fermion rest mass $E$ at
$\alpha_c$ is necessarily the same for the $0^-$ and $0^+$ states
(certainly, such is not the case in the one-body limit). Note, also that the
result, $\alpha_c = 2/\sqrt{3}$ for $0^\pm$ states is independent of the
masses, that is we expect it to be the same for all finite $m_1/m_2$.  The
value $\alpha_c = 2/\sqrt{3}$ is different, and somewhat larger, than the
known one-body limit (Dirac-Coulomb) value of $\alpha_c= 1$ for $|\kappa|
=1$ states. Also, this value is much smaller that the value $\alpha_c = 2$,
which is obtained for the two-fermion Coulomb QED case  (where $\xi=\eta=0$)
for $0^\pm$ states.

For the Coulomb potential $\ds V= -{\alpha \over r}$, where
$\ds \alpha = {{|q_1q_2|}\over {4\pi}}$, it is often convenient
to rescale the radial variable, that is to let $\rho =  r/a$,
where $a$ is a suitable scale parameter. For example, the radial functions
$s$, $t$, $u$, $v$ have the large $r$ (negligible $V$ and $K/r$)
behaviour $s \sim e^{-\rho}$, etc., for positive energy $J=0$ bound states,
where $a$ is given by
%
\begin{equation} \label{7-5}
{1\over {a^2}} =  {{[m_+^2-E^2][E^2 - m_-^2]}\over {4E^2}} \qquad
{\rm or} \qquad {1\over {a^2}} =  m^2 - \left({E\over 2}\right)^2 \quad {\rm
if} \quad m_1=m_2 .
\end{equation}
Eq. \re{7-5} implies that $a$ is positive only for $|m_1-m_2| \le E \le
m_1+m_2$, which means that the bound state spectrum must lie in this domain
({\sl c.f.} \re{7-4a}). From this, and in analogy with the scalar model
results, we can infer that the critical value $E(\alpha_c = 2 / \sqrt{3})$
lies between $E=m_+$ and $E=|m_-|$, and likely closer to the former rather
than the latter.

With the rescaling $\rho =  r/a$, equations \re{7-1}-\re{7-4} become
modified slightly, in that $r$ is replaced by $a\rho$ in all of them.  For
purposes of numerical integration of the radial equations the scale
parameter $a$ can be chosen to be anything that is convenient, be it
that given in Eq. \re{7-5}, or $a= 1$ or $\ds a= {1\over{\mu \alpha}}$,
or whatever.

For a power series analysis of the radial equations it is useful
to make the replacement $s= \overs e^{-\rho}$, etc. Assuming
solutions of the form
%
\begin{equation} \label{7-6}
\overs =  \rho^\gamma [a_0 + a_1\rho + a_2\rho^2 + \cdots],
\end{equation}
\begin{equation} \label{7-7}
\overt =  \rho^\gamma [b_0 + b_1\rho + b_2\rho^2 + \cdots],
\end{equation}
\begin{equation} \label{7-8}
\overu =  \rho^\gamma [c_0 + c_1\rho + c_2\rho^2 + \cdots],
\end{equation}
\begin{equation} \label{7-9}
\overv =  \rho^\gamma [d_0 + d_1\rho + d_2\rho^2 + \cdots],
\end{equation}
we find , upon substitution into the radial equations for
$\overs$, $\overt$, $\overu$, $\overv$ and equating coefficients
of powers of $\rho^{\gamma +\nu -1}$, that the coefficients
$a_j, b_j, c_j, d_j$ must satisfy the following recursion relations:
%
\begin{equation} \label{7-10}
a(m_+ - E)a_{\nu-1} - \alpha a_{\nu} - (\gamma+K+\nu)b_{\nu}
+\delta b_{\nu-1} - (\gamma+K+\nu)c_{\nu} +\delta c_{\nu-1}
- \xi \alpha d_\nu =  0,
\end{equation}
\begin{equation} \label{7-11}
(\gamma-K+\nu)a_{\nu} -\delta a_{\nu-1} + a(m_- -E)b_{\nu-1}
- \alpha b_{\nu} + (\gamma-K+\nu)d_{\nu} -\delta d_{\nu-1}
- \eta \alpha c_\nu =  0,
\end{equation}
\begin{equation} \label{7-12}
(\gamma-K+\nu)a_{\nu} -\delta a_{\nu-1} +a(- m_- -E)c_{\nu-1} -
\alpha c_{\nu} + (\gamma-K+\nu)d_{\nu} -\delta d_{\nu-1}
- \eta \alpha b_\nu =  0,
\end{equation}
\begin{equation} \label{7-13}
(\gamma+K+\nu)b_{\nu} -\delta b_{\nu-1} + (\gamma+K+\nu)c_{\nu}
-\delta c_{\nu-1} + a(m_+ + E)d_{\nu-1} + \alpha d_{\nu}
+ \xi \alpha a_\nu =  0,
\end{equation}
where $\delta =  1$. If $\delta =  0$ then \re{7-10}-\re{7-13}
are the recursion relations for the power series representations of the
functions $s(r)$, etc., rather than for $\overs (r)$, etc.

For $\nu = 0$, and bearing in mind that
$a_{-1} = b_{-1} = c_{-1} = d_{-1} = 0$, \re{7-10}-\re{7-13} yield four
coupled homogeneous equations for the parameters $a_0$, $b_0$,
$c_0$, $d_0$, which have non-trivial (and non-singular) solutions only if
%
\begin{equation} \label{7-14}
\gamma =  \sqrt{ K^2- {1 \over 4} (1+\xi)(1+\eta)\alpha^2    }
=  \sqrt{1 - {{3\alpha^2} \over 4} }
\end{equation}
for the $J=0$ states, for any values of $m_1$, $m_2$, whereupon
%
\begin{equation} \label{7-15}
{d_0 \over a_0} =  1, \quad {b_0 \over a_0} =
{c_0 \over a_0} =  -{ {(1+\xi)\alpha} \over {2(\gamma +K)}}
=  {2(\gamma-K)\over {(1+\eta)\alpha} } .
\end{equation}

The condition \re{7-14} implies that the radial equations have
real bound state solutions of the form \re{7-6}-\re{7-9} only for
$\alpha \leq {2 \over \sqrt{3}}$, for any values of $m_1$ and $m_2$.
This, in turn, implies that $\alpha_c \le {2 \over \sqrt{3}}$ for the
$0^{\mp}$ states for any (finite) values of $m_1$ and $m_2$, in agreement
with the ultra relativistic limit discussed above.
This condition for bound states, $\alpha \le {2 \over \sqrt{3}}$,
is additional to the one that follows from Eq. \re{7-5}, namely
that $|m_1-m_2| \le E \le m_1+m_2$.

The recursion relations \re{7-10}-\re{7-13}, with \re{7-14} and \re{7-15},
can be used to generate the power series form of the solutions of Eqs.
\re{7-6}-\re{7-9}. These series converge in the domain $r{^{~<} \!\!\!{_
\sim}} \; \alpha / m_+$, as discussed below and in \cite{DdiL96}. Such a
series can be used, for example, as a starting procedure for the numerical
integration of the radial equations \re{7-1}-\re{7-4}.

Unlike in the Dirac case, the recursion relations \re{7-10}-\re{7-13} do
not admit power series solutions of the form \re{7-6}-\re{7-9}, which
terminate at the same power, say $\nu= n^\prime$, so that
$a_{n^{\prime}+1}= b_{n^{\prime}+1}= c_{n^{\prime}+1}=
d_{n^{\prime}+1}= 0$. In particular, the ground state
solution is not of the simple form
%
\begin{equation} \label{7-16}
\overs= a_0\rho^\gamma,\;\;\overt= b_0\rho^\gamma,\;\;\overu=
c_0\rho^\gamma,\;\;\overv= d_0\rho^\gamma
\end{equation}
as it is for the two radial Dirac equations. This is perhaps to
be expected, since in the Dirac case there are only two functions,
say $\overs$ and $\overt$, and four unknowns to be determined,
namely $\ds {b_0 \over a_0}$, $\gamma$, $a$ and $E$.
Since the two coupled radial Dirac equations yield four equations
(the coefficients of $\rho^\gamma$ and of $\rho^{\gamma-1}$),
it is not surprising that a solution is obtained.
In the present case, we have four coupled radial equations
\re{7-1}-\re{7-4}, which yield eight equations (the coefficients of
$\rho^\gamma$ and of $\rho^{\gamma-1}$) to be satisfied by the
six unknowns of the proposed solutions \re{7-16}, namely $b_0$, $c_0$,
$d_0$, $\gamma$, $a$ and $E$. Thus the system is overdetermined
and no solution of the form \re{7-16} is possible. This situation
persists for any solution of the form \re{7-6}-\re{7-9}
where the polynomials all terminate at the same degree.
Therefore, we shall solve the radial equations \re{7-1}-\re{7-4} numerically.

Equations \re{7-1} - \re{7-4} are not independent. Indeed,  elementary
manipulations of these equations, namely subtracting Eq. \re{7-4} from Eq.
\re{7-1} and similarly Eq. \re{7-3} from Eq. \re{7-2} show that
%
\be   \label{7-6a}
v(r) = {{E-m_+  - (1-\xi) V(r)}\over{E+m_+  -(1-\xi)V(r)}} s(r), \;\;\;
u(r) = {{E-m_- - (1-\eta)V(r)}\over{E+m_-  - (1-\eta)V(r)}} t(r).
\ee
Thus, the number of equations can be reduced from four to two.

We introduce the auxiliary functions $f(r) = s(r) + v(r)$ and $g(r) = t(r) +
u(r)$. Then, adding Eqs. \re{7-1} to Eq. \re{7-4} and Eq. \re{7-2} to Eq.
\re{7-3}, and using Eq. \re{7-6a} yields the equations
%
\be \label{7-7a}
f^{\prime} (r) = {K \over r} f(r) + W_g (r) g(r),  \qquad  -g^{\prime}  (r)
= {K \over r} g(r) + W_f (r) f(r),
\ee
where
%
\be \label{7-8a}
W_g (r) =  {1 \over 2} \left[ E - V_\eta (r) - {{(m_1-m_2)^2} \over
{E-{\overline V}_\eta (r) } } \right] ,
\ee
\be \label{7-9a}
W_f (r) =  {1 \over 2} \left[ E - V_\xi (r) - {{(m_1+m_2)^2} \over
{E-{\overline V}_\xi (r) } } \right] ,
\ee
and where
%
\begin{eqnarray}
V_\xi (r) = (1+\xi) V(r), \qquad  {\overline V}_\xi (r) = (1-\xi) V(r),
\nonumber \\
V_\eta (r) = (1 + \eta) V(r), \qquad {\overline V}_\eta (r)  =
(1 - \eta) V(r). \label{7-10a}
\end{eqnarray}

For the present QED case in the Coulomb gauge, for the $0^-$ states (for
which $\xi=2, \eta=0$), $V_\eta= {\overline   V}_\eta=V$ while $V_\xi= 3V$
and ${\overline V}_\xi= -V$.  In this case we see that $W_f (r)$ is
singular at $r_1 = \alpha / E = \alpha^2  (m_r / E) (1/m_r \alpha)$,  where
$m_r$ is the reduced mass and $1 / m_r \alpha$ is the reduced Bohr  radius.
This singular point is quite close to the origin (in units of the  reduced
Bohr radius) for small $\alpha$.  The appearance of this  singularity may
signal difficulties in the numerical determination of  eigensolutions of the
equations \re{7-7a} by standard ``shooting"  methods. For the $0^+ (^3P_0)$
states (for which $\xi=0, \eta=2$), the  singularity at $r_1= \alpha/E$
occurs in $W_g$, but only if $m_1 \ne  m_2$.  Thus for the equal-mass $0^+$
states, equations \re{7-7a} have  only the usual $1/r$ singularities at the
origin, and are amenable to  solution by standard methods, as discussed
below.


\section{Perturbative determination of the relativistic correction to
the two-body eigenenergies for $J=0$ states}

\setcounter{equation}{0}
\renewcommand{\theequation}{8-\arabic{equation}}

Equations \re{7-7a}  can be written in the matrix form
%
\be \label{8-1}
H |\psi \rangle = \epsilon |\psi \rangle \qquad {\rm where} \qquad
H = \left[\matrix{\epsilon - W_f&- {d \over {dr}} - {K \over r}\cr
                {d \over {dr}} - {K \over r}&\epsilon - W_g\cr}\right],
\qquad \psi =   \left[\matrix{f  \cr g \cr}\right],
\ee
and where $\epsilon = E - (m_1+m_2)$.
If $W_f $ is replaced by $W_f^{\rm nr} = \epsilon_{\rm nr} - V$ and $W_g$
by $W_g^{\rm nr} = 2 \mu$, where $\mu = m_1 m_2 /(m_1+m_2)$, then Eq.
\re{7-7a}, or \re{8-1}, is  equivalent to the radial Schr\"odinger
equation. The first-order  correction to the non relativistic energy
$\epsilon_{\rm nr} =  - {1  \over 2} \mu \alpha^2 {1 \over n^2}$ is then
given by
%
\be \label{8-2}
\Delta \epsilon =  {{\langle \psi _{\rm nr} | H- H_{\rm nr}| \psi_{\rm nr}
\rangle } \over  {\langle \psi _{\rm nr} | \psi_{\rm nr} \rangle }} =  {
{{\langle f _{\rm nr} | \epsilon - \epsilon_{\rm nr} + W_f ^{\rm nr} - W_f
| f_{\rm nr} \rangle } + {\langle g _{\rm nr} |  \epsilon - \epsilon_{\rm
nr} + W_g ^{\rm nr} - W_g | g_{\rm nr} \rangle }}
 \over  {{\langle f _{\rm nr} |  f_{\rm nr} \rangle } +
{\langle g _{\rm nr} |  g_{\rm nr} \rangle }}    }
\ee
If we expand the coefficients $W_f$ and $W_g$ (Eqs. \re{7-8} and \re{7-9})
in powers of $V/m_i$, and keep only the lowest-order terms, we obtain
%
\be \label{8-2a}
 \epsilon - \epsilon_{\rm nr} + W_f ^{\rm nr} - W_f  \simeq {1 \over
{2m_+}} (\epsilon_{\rm nr} - V_\xi)^2
\ee
\be  \label{8-2b}
 \epsilon - \epsilon_{\rm nr} + W_g ^{\rm nr} - W_g  \simeq   -
\left( 1   - 2{ \mu \over {m_+}} \right) \epsilon _{\rm nr}  +
{1 \over 2} V_\eta  +   \delta^2 \, {\overline V}_\eta ,
\ee
where $\delta = m_- / m_+$.
This leads to the following $O(\alpha^4)$ correction to the
non-relativistic energy for the $J=0$ states:
%
\begin{eqnarray}
\Delta \epsilon  = {1 \over { 2m_+}} \left[ \epsilon_{\rm nr}^2 - 2
\epsilon_{\rm nr}\, (1-\xi) \langle V \rangle + (1-\xi)^2 \langle V^2
\rangle \right ]  \nonumber   \\
- \left( 1 - 2{ \mu \over {m_+}} \right) \epsilon _{\rm nr} \, \langle
\langle 1 \rangle \rangle  +{1 \over   2}\left( (1 +\eta) + (1 - \eta)
\delta^2 \right)  \langle \langle V   \rangle \rangle .
\label{8-3}
\end{eqnarray}
We use the notation $ \langle X \rangle = \langle f_{\rm nr} | X | f_{\rm
nr}  \rangle /   \langle f_{\rm nr} |  f_{\rm nr}  \rangle $
but $ \langle \langle X  \rangle \rangle =
\langle g_{\rm nr} | X | g_{\rm nr}  \rangle /   \langle f_{\rm nr} |
f_{\rm nr}  \rangle $.

For the  $0^- \, (n \; ^1S_0)$ states  (for which
$K=1$ and $ \xi=2,  \eta=0$) this formula gives for the lowest order
relativistic correction  the result
%
\be \label{8-4}
\Delta \epsilon \left( 0^- \, (n \; ^1S_0) \right) =  \mu \alpha^4
\left\{ {1 \over n^4} \left( {3 \over 8} - {1 \over 8} {\mu \over m_+}
\right) + {1 \over n^3} \left( -{1 \over 2} +2 {\mu \over m_+} \right)
\right\}\ee
which becomes $\ds {{11} \over {64}} m \alpha^4 $ in the equal
mass $m_1=m_2=m$ case. This does not agree with the known
Positronium value   of $\ds - {{21} \over {64}} m \alpha^4 $ \cite{BS57}.
This is not surprising, since the Breit equation, without modification,
is known to give the incorrect fine structure for Hydrogen and positronium.
Brown and Ravenhall \cite{BR51} argue that the reason for this,
is the mixing of positive and negative energy one-particle states (which
arises, in our formalism, because of our use of the ``empty'' vacuum
\re{3.4}). This difficulty of the unmodified Breit equation  is discussed
in various works ({\sl e.g.} Refs. \cite{BR51,Chrap53,BS57}). The
modification that is needed to bring the result into agreement with the
observed fine structure of H or Ps is to subtract off  the expectation
value of the operator \cite{Chrap53,Bar55}
%
\be \label{8-7}
H^\prime =   {\alpha^2 \over { 4 m_+ r^2}} \left( 3 - 2 \arrowsig_1 \cdot
\arrowsig_2  + \sigma_{1r} \sigma_{2r} \right) \qquad \sigma_{ir} =
\arrowsig_i \cdot \br / r ,
\ee
where, in this equation, we use the notation of \cite{BS57}. The
expectation value of \re{8-7} (with respect to the non-relativistic
eigenfunctions) is
%
\be \label{8-8}
 \bra H^\prime \ket_{\rm nr} = \mu \alpha^4 {{1 -\delta^2}
\over {n^3   (2J+1)} } f_S
 =  \mu \alpha^4 {{4 f_S} \over {n^3 (2J+1)} } {\mu \over m_+} ,
\ee
where $f_S = 1$ for the singlet $S=0$ states, while
$f_S = 1/4$ for the triplet $S=1$ states with $J >0$, but $f_S =0$ for the
triplet states with $J=0$ (see, also, section 10 below).  The expression
\re{8-8} gives the value $\ds {1 \over 2} m \alpha^4$ for the equal-mass
ground state, which, when subtracted from the `Breit' value of  $\ds {{11}
\over {64}} m \alpha^4$ gives the expected positronium result $\ds  - {{21}
\over {64}} m \alpha^4 $. More generally for arbitrary masses, if we
subtract \re{8-8} from the expression \re{8-3}, we obtain the ``corrected"
result
%
\be \label{8-9}
\Delta \epsilon_c \left(n \; 0^- \,(^1S_0) \right) =  \mu \alpha^4
\left\{ {1 \over n^4} \left( {3 \over 8} - {1 \over 8} {\mu \over m_+}
\right) + {1 \over n^3} \left( -{1 \over 2} - 2 {\mu \over m_+} \right)
\right\} .
\ee
This same result \re{8-9} was obtained previously for the $n=2$ state by
Darewych and Horbatsch \cite{DH}b, who used a perturbative approximation on
variationally derived equations.

Somewhat surprisingly, the unequal-mass $O(\alpha^4)$ corrections
for arbitrary states seem to have been worked out fully only relatively
recently. We refer to the work of Connell \cite{Connell}, who used a
quasipotential formalism based on the work of Todorov \cite{Tod71},  and of
Hersbach \cite{Hersb}, who used a formalism based on a
relativistic generalization of the Lippmann-Schwinger equation due to De
Groot and Ruijgrok \cite{DeGR}. Our corrected expression \re{8-9} agrees
with the results of these authors. (The $O(\alpha^4)$ corrections for
Hydrogen and muonium quoted in standard references are expansions in
$m_1/m_2$ ({\sl e.g.} \cite{SY}, \cite{Mohr}.))

For the  $n \, 0^+ \, (^3P_0)$ states  (for which $K=-1$ and  $ \xi=0,
\eta=2$) equation \re{8-3} gives
%
\be \label{8-5}
\Delta \epsilon \left(n \; 0^+ \, ( ^3P_0) \right) = \mu \alpha^4
\left\{ {1 \over n^4} \left( {3 \over 8} - {1 \over 8}
{\mu \over m_+} \right) + {1 \over n^3} \left( - {1 \over 2} - {2 \over 3}
{\mu \over m_+} \right)  \right\} ,
\ee
which {\sl does} agree, in the equal mass case, with the Ps values
for all the $n \;  0^+ (^3P_0)$ states, as well as with the unequal-mass
expressions of Connell \cite{Connell} and Hersbach \cite{Hersb} for these
states. This agreement implies that the ``correction" $\bra H^\prime
\ket_{\rm nr}$ vanishes for the $^3P_0$ states, as indeed it does.

We might note, in passing, that formula \re{8-3}  gives the correct
$O(\alpha^4)$ results for the Coulomb QED ($ \xi=\eta=0$)  case
\cite{DdiL96}, for which the $W$ coefficients are non-singular for $r > 0$.


\setcounter{equation}{0}
\renewcommand{\theequation}{9-\arabic{equation}}

\section{Numerical solutions for some $J=0^+$ states}

In the case of equal masses $\;m_1=m_2\equiv m\;$ the radial equations
\re{7-7a} for $J=0^+ (^3P_0)$ states are free of singularities. Thus the
boundary value problem is well posed, and it can be solved by means of a
standard numerical ODE-solving procedure. We solved it by the ``shooting"
method using the Maple Runge-Kutta programme.

The corresponding perturbative spectrum ({\sl c.f.} \re{8-5}),
%
\begin{equation}\lab{N1}
E/m = 2 - \frac{\alpha^2}{4n^2} + \alpha^4
\left\{\frac{11}{64n^4} -
\frac{1}{3n^3}\right\},
\end{equation}
agrees with the orthopositronium spectrum since the contribution of
the extra terms
({\sl c.f.} \re{8-7}) (caused by positive-negative energy mixing)
vanishes in this case.

In the Table 1 the numeric and perturbative results are presented for the
lowest-energy $0^+$ state  ({\sl i.e.}, $J=0$, $\ell=1$, $n=2$) for
different values of $\alpha \le \alpha_c=2/\sqrt3\approx1.1547005383792515$.
%
%
\begin{table}[t]
\noindent
\caption{
Values of $E/m$ for the $n=2$, $0^+(^3P_0)$ state ($m_1=m_2\equiv m$).
}
\noindent

\begin{center}
\begin{tabular}{|l|l|l|}
\hline
\hspace{1ex}$\alpha$&
\hspace{1ex}Perturbative &
\hspace{2ex}Numeric\\
&(equation \re{N1})&
(equations \re{7-7a})\\
\hline

1/137 & 1.999 996 669 9532 & 1.999 996 669 9532 \\
0.01    & 1.999 993 749 6908 & 1.999 993 749 6908 \\
0.05    & 1.999 843 556 7220 & 1.999 843 5564 \\
0.1     & 1.999 371 907 5521 & 1.999 371 886 \\
0.5   & 1.982 442 220 0521 & 1.982 028 02 \\
0.7   & 1.961 950 032 5521 & 1.957 997 74 \\
0.9   & 1.929 085 449 2188 & 1.902 4531  \\
1.0   & 1.906 575 520 8333 & 1.838 781 05\\
1.1   & 1.879 098 470 0521 & 1.688 2317  \\
1.15  & 1.863 256 642 6595 & 1.436 9434  \\
1.154 & 1.861 924 191 0995 & 1.355 170 76\\
1.1547 & 1.861 689 995 0549 & 1.301 3199  \\
1.1547005383792 & 1.861 689 814 8148 & 1.299 74 \\
\hline
\end{tabular}
\end{center}
\end{table}

Note that the perturbative $O(\alpha^4)$ results are unreliable for
$\alpha > 0.5$.
Our numerical results suggest that, for this $n=2$, $0^+$ state,
$E(\alpha_c = 2/\sqrt{3}) \simeq 1.29974 \; m $, which is smaller
than the scalar theory value $E (\alpha_c=2)/m = \sqrt{2} = 1.4142..$ (see
below \re{7-4a}).

There are, as we explained in section 7, $E > 0$ ``mixed
energy" solutions of the form $E/m =  \alpha + O(\alpha^3)$, which are
unphysical, because they do not have the Balmer non-relativistic limit. We
do not list such solutions here, though they can be calculated readily
enough in the same way as those in the Table 1. This unphysical branch of
the $n=2\,$  $0^+$ state rises uniformly from zero at $\alpha =0$ to join
the physical branch of Table 1 smoothly at $E(\alpha_c)$. As mentioned
previously, the two branches together resemble a distorted semicircle ({\sl
c.f.} Eq. \re{7-4a}).

Analogous results for the $n=3$ and $n=4$ $0^+ (^3P_0)$ equal-mass
two-fermion energies are given in Table 2.  The qualitative behaviour of
$E(\alpha)/m $ for these states is similar to that for the lowest such
state ($n=2$), except that the critical value of $\alpha$ increases with
$n$, as it does in the case of the analytically solvable scalar model of
Eq. \re{7-4a}. However, here we obtain $\alpha_c / n = 0.64987, \,
0.6024745, \, 0.4871785 \,$ for the $n=2,3$ and $4$ states respectively, in
contrast to the scalar model values $\alpha_c / n = 1$ for all $n$.

\begin{table}[t]
\noindent
\caption{
Values of $E/m$ for the $n=3,4 \;\, 0^+(^3P_0)$ states ($m_1=m_2\equiv m$).
}
\noindent

\begin{center}
\begin{tabular}{|l|l|l|l|}
\hline
\hspace{1ex}$n$&
\hspace{1ex}$\alpha$&
\hspace{1ex}Perturbative &
\hspace{2ex}Numeric\\
&&(equation \re{N1})&
(equations \re{7-7a})\\
\hline
3&1/137 & 1.999 998 519 9892 & 1.999 998 519 9892  \\
&0.01    & 1.999 997 222 1200 & 1.999 997 222 1200 \\
&0.05    & 1.999 930 491 6570 & 1.999 930 491 6570 \\
&0.1     & 1.999 721 199 8457 & 1.999 721 193 728 \\
&0.5   & 1.992 416 570 2160 & 1.992 303 736 \\
&0.7   & 1.983 934 162 8086 & 1.982 908 59 \\
&0.9   & 1.970 792 187 5000 & 1.964 6458 \\
&1.0   & 1.961 998 456 7901 & 1.948 028 09 \\
&1.1   & 1.951 420 273 9198 & 1.918 6196 \\
&1.15  & 1.945 382 459 2496 & 1.884 4358 \\
&1.154 & 1.944 876 373 1198 & 1.875 613 \\
&1.1547 & 1.944 787 448 4071 & 1.870 5697 \\
&1.1547005383792 & 1.944 787 379 9726 & 1.870 4234 \\
\hline
4&1/137 & 1.999 999 167 4974 & 1.999 999 167 4974 \\
&0.01    & 1.999 998 437 4546 & 1.999 998 437 4546 \\
&0.05    & 1.999 960 909 1441 & 1.999 960 909 1441 \\
&0.1     & 1.999 843 296 3053 & 1.999 843 2939 \\
&0.5   & 1.995 810 190 8366 & 1.995 766 2947\\
&0.7   & 1.991 254 429 1178 & 1.990 8649 \\
&0.9   & 1.984 367 059 3262 & 1.982 1441 \\
&1.0   & 1.979 838 053 3854 & 1.974 9816 \\
&1.1   & 1.974 451 206 4616 & 1.963 685 \\
&1.15  & 1.971 400 789 5152 & 1.952 562 \\
&1.154 & 1.971 145 810 1087 & 1.950 1225 \\
&1.1547 & 1.971 101 018 2659 & 1.948 751 \\
&1.1547005383792 & 1.971 100 983 7963 & 1.948 714 \\
\hline
\end{tabular}
\end{center}
\end{table}

Note that the critical value of the two-body mass, $E(\alpha_c) / m$,
increases with $n$, in contrast to the scalar model, for which
$E(\alpha_c) / m = \sqrt{2}$ for all $n$.

 	Figure 1 is a plot of the unnormalized reduced radial
wave functions  $s(r)$,	$t(r)=u(r)$ and $v(r)$ (see Eqns.
\re{7-1}-\re{7-4}) in the case
of equal massless, $m_1=m_2\equiv m$,  for the lowest-energy
$\;n=2\;0^+(^3{\rm P}_0)$ states, when $\alpha=1$. These  wave functions
are qualitatively similar to those obtained for these states
in the Coulomb QED case \cite{DdiL96}. The ``large'' component $s(r)$
is  nodeless while the ``small'' components $t(r)=u(r)$ and the
``doubly small'' one  $v(r)$ have one node. The node at the origin,
$r=0$, is a consequence of our use of reduced radial wave functions
$s(r)$, etc., rather than the actual $s(r)/r$, etc. Indeed, the
wave functions behave at small $r$ as follows:
%
\begin{equation}\lab{N2}
s(r) \approx v(r) \approx {\rm const}\cdot\alpha r^\gamma, \qquad
t(r) = u(r) \approx {\rm const}\cdot(2/3)(\gamma+1)r^\gamma,
\end{equation}
where $0<\gamma<1$ (see Eq. \re{7-16}). Thus the
matrix wave function $F(\B r)$ is singular, $F(\B r)\sim r^{\gamma-1}$,
as happens also in the one-body Dirac equation with a Coulomb potential.
Nevertheless, $F(\B r)$ is normalizable for all $\alpha$ up to and
including  $\alpha=\alpha_c$, at which point
$t(0)=u(0)=(2/3)s(0)=(2/3)v(0)\ne0$, as can be seen in Figures 2 and 3.

	Figure 3 represents the excited $\;n=3\;0^+(^3{\rm P}_0)$  state
for the critical coupling strength $\alpha_c = 2/\sqrt{3}$.
In this case, the wavefunction $s(r)$ has one node while $t(r)$ and
$v(r)$ have two nodes. This behaviour differs from  that found in CQED
\cite{DdiL96} for the same case (where the number  of nodes was two,
one and three, respectively). In the $\;n=4\;0^+(^3{\rm  P}_0)$ case
(Figure 4) each of the functions $s(r)$, $t(r)$ and $v(r)$
gets one more node . This tendency likely continues
for higher values of the quantum number $n$.

\begin{figure}[p]
\epsfxsize=160mm
\epsffile{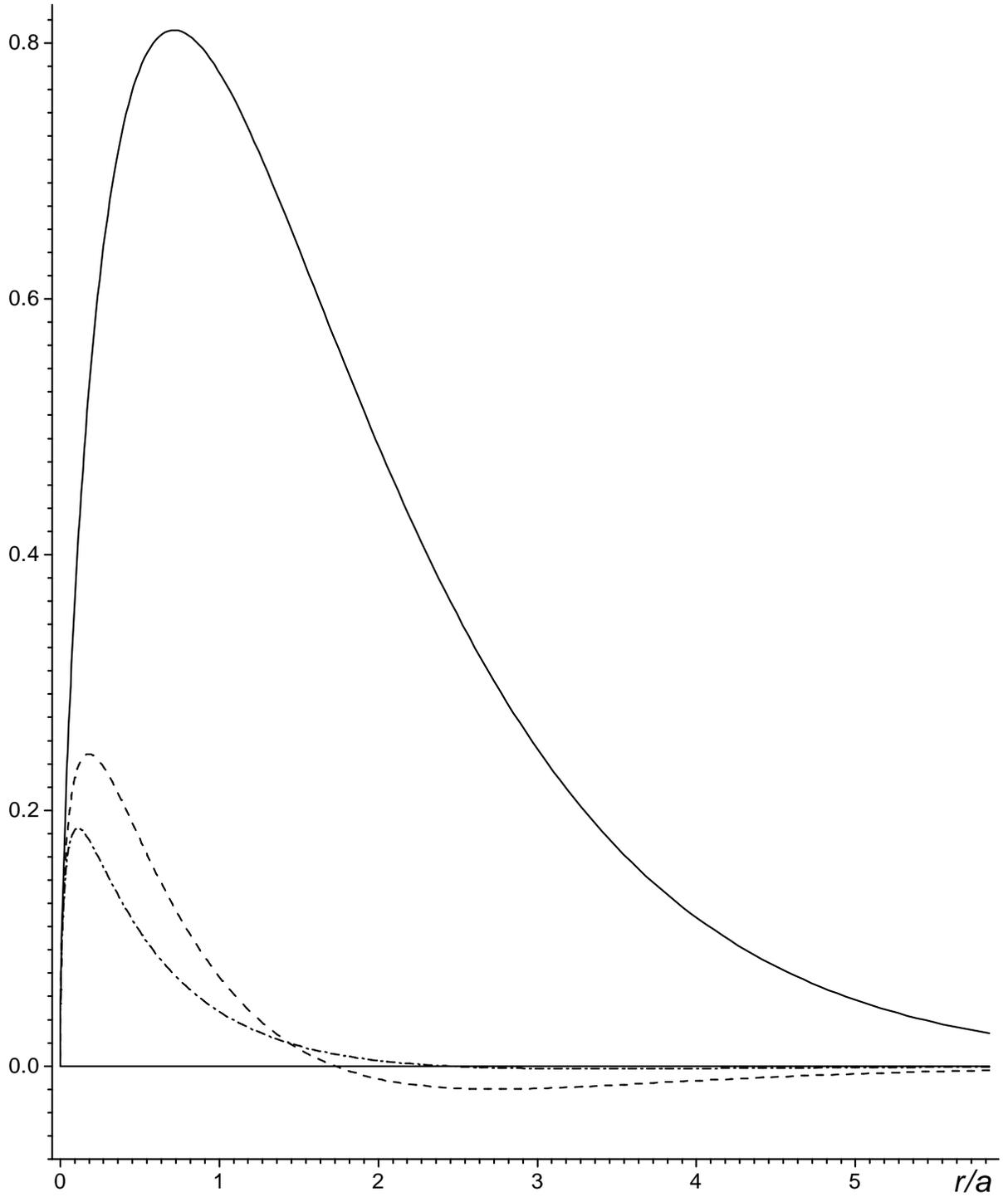}
\caption{
Reduced radial wavefunctions for the lowest $m_1=m_2\equiv m$\ \
$\;n=2\;0^+(^3{\rm P}_0)$ state for $\alpha=1$, $E/m=1.838781$.
$s(\rho)$: full curve; $t(\rho)=u(\rho)$: broken curve; $v(\rho)$:
chain  curve. $\rho=r/a$, where $a=2.542291(1/m)$.
}
\end{figure}

\begin{figure}[p]
\epsfxsize=160mm
\epsffile{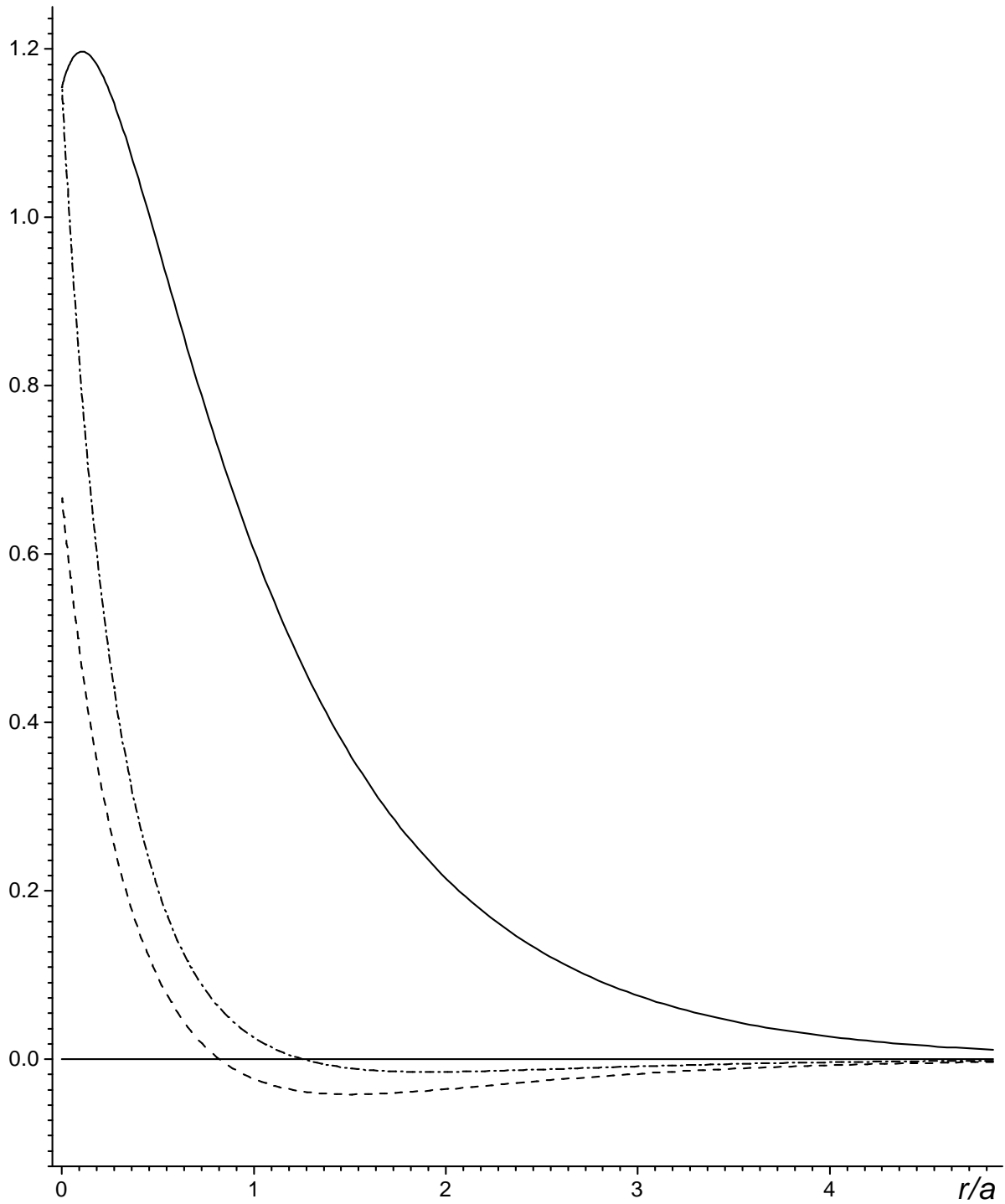}
\caption{
Same as Figure 1 but $\alpha=\alpha_c=2/\protect\sqrt{3}$, $E/m=1.29974$
and $a=1.315711(1/m)$.
}
\end{figure}

\begin{figure}[p]
\epsfxsize=160mm
\epsffile{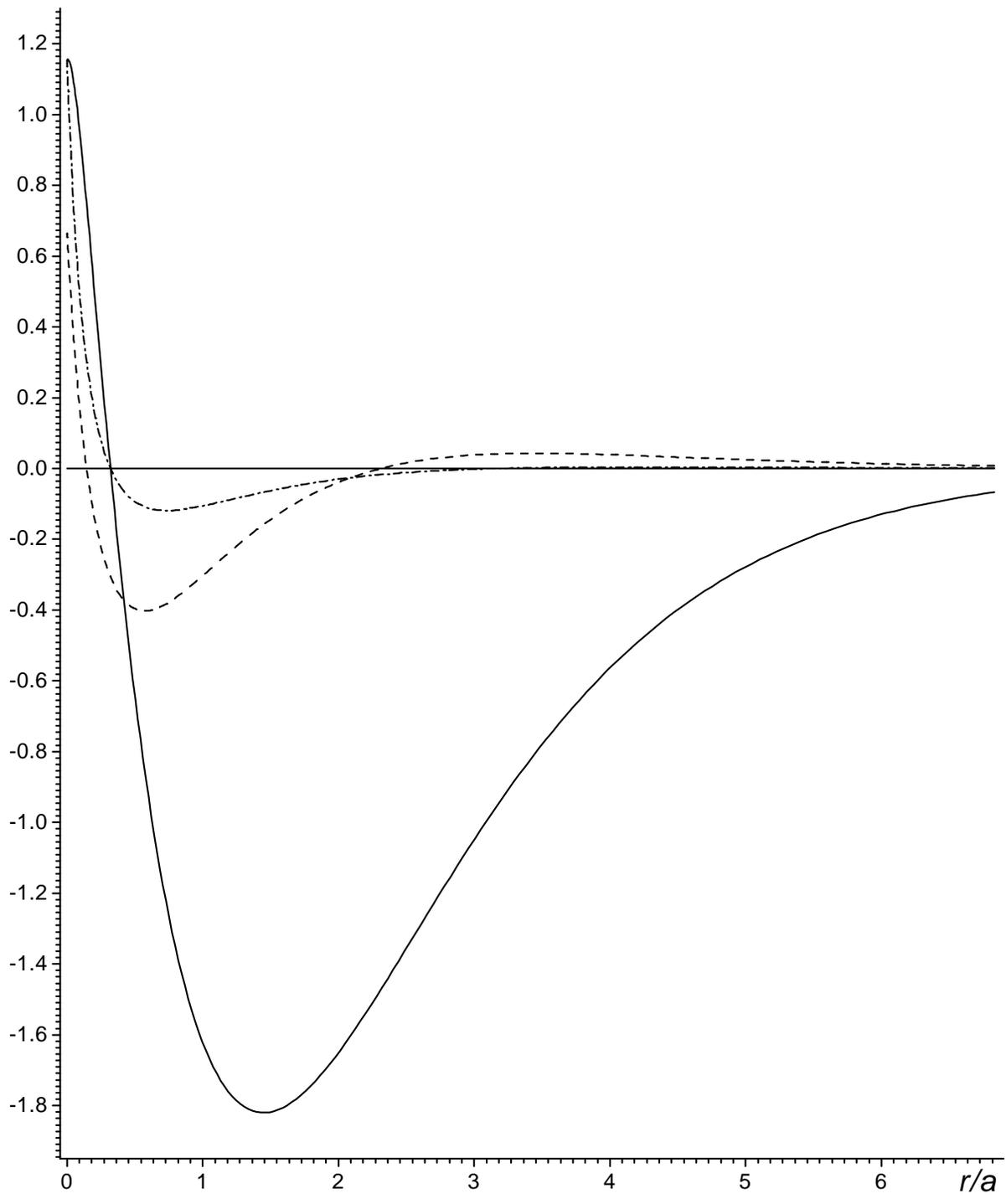}
\caption{
Same as Figure 1 but for the excited $\;n=3\;0^+(^3{\rm P}_0)$  state,
with $\alpha=\alpha_c=2/\protect\sqrt{3}$, $E/m=1.870423$ and
$a=2.824148(1/m)$. }
\end{figure}

\begin{figure}[p]
\epsfxsize=160mm
\epsffile{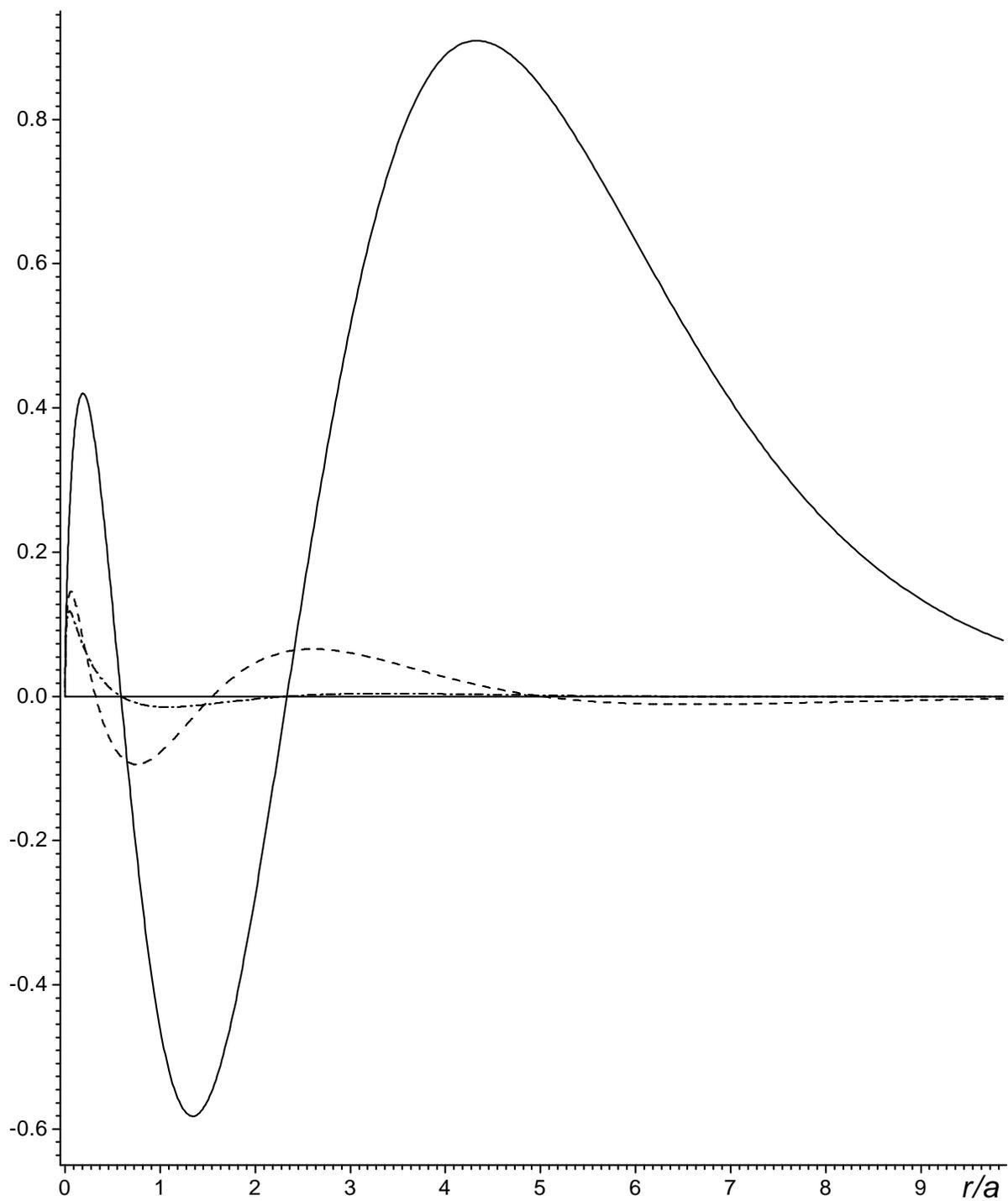}
\caption{
Same as Figure 1 but for the excited $\;n=4\;0^+(^3{\rm P}_0)$
state,  with $\alpha=1$, $E/m=1.974982$ and $a=6.342094(1/m)$.
}
\end{figure}

\vfill \eject


\setcounter{equation}{0}
\renewcommand{\theequation}{10-\arabic{equation}}

\section{Radial reduction for $J>0$ states}  

For states with $J > 0$, the eigenstate problem  reduces to a set of eight
first-order differential equations for the functions $s_1(r)\dots v_2(r)$
and the energy $E$ ({\sl c.f.} Eqns. \re{6-1} and \re{6-2}). It is
convenient to present this set in the following matrix form.
Let us introduce the 8-dimensional vector-function:
%
\begin{equation}\lab{R3}
\s X(r) = \left[
\begin{array}{c}
s_1(r) \\ s_2(r) \\ t_1(r) \\ \vdots \\ v_2(r)
\end{array}
\right].
\end{equation}
Then the set of radial equations reads:
%
\begin{equation}\lab{R4}
{\cal H}\s X(r) \equiv \left\{\s H \frac{d}{dr} + \s U(r)\right\}\s X(r) =
E\s X(r),
\end{equation}
where the 8$\times$8 matrix $\s U(r)$ has the following structure:
%
\begin{equation}\lab{R5}
\s U(r) = \s M +[\s G - \alpha(\s I + \s S)]/r.
\end{equation}
Here $\s I$ is the unit matrix, $\s M$ is diagonal,
%
\begin{equation}\lab{R6}
\s M = \left[\mbox{\scriptsize$
\begin{array}{rrrrrrrr}
\lefteqn{\ m_+}&&&&&&&\\
&\lefteqn{\ m_+}&&&&&&\\
&&\lefteqn{\ m_-}&&&&\smash{\lefteqn{\hspace{1em}\mbox{\Huge 0}}}&\\
&&&\lefteqn{\ m_-}&&&&\\
&&&&\lefteqn{-m_-}&&&\\
&&&&&\lefteqn{-m_-}&&\\
&\smash{\lefteqn{\mbox{\Huge 0}}}&&&&&\lefteqn{-m_+}&\\
&&&&&&&-m_+
\end{array}$}
\right],
\end{equation}
$m_\pm = m_1 \pm m_2$, and the form of 8$\times$8 matrices $\s H$,
$\s G$, and $\s S$ depends on the parity $P$:
%
\begin{eqnarray}
\s H &=& \left[\mbox{\scriptsize$
\begin{array}{cccccccc}
 0 &  0 &  -A & B &  -A & B &  0 &  0\\
 0 &  0 & B & A &  -B &  -A &  0 &  0\\
 A & -B &  0 &  0 &  0 &  0 &  A &  B\\
-B & -A &  0 &  0 &  0 &  0 & -B &  A\\
 A &  B &  0 &  0 &  0 &  0 &  A & -B\\
-B &  A &  0 &  0 &  0 &  0 & -B & -A\\
 0 &  0 &  -A & B &  -A & B &  0 &  0\\
 0 &  0 &  -B &  -A & B & A &  0 &  0
\end{array}$}
\right],\lab{R7}\\
\s G &=& \left[\mbox{\scriptsize$
\begin{array}{cccccccc}
 0 &  0 &  -(J{+}1)A &  -JB &  -(J{+}1)A &  -JB &  0 &  0\\
 0 &  0 & (J{+}1)B &  -JA &  -(J{+}1)B & JA &  0 &  0\\
-(J{+}1)A & (J{+}1)B &  0 &  0 &  0 &  0 & -(J{+}1)A & -(J{+}1)B\\
-JB & -JA &  0 &  0 &  0 &  0 & -JB &  JA\\
-(J{+}1)A & -(J{+}1)B &  0 &  0 &  0 &  0 & -(J{+}1)A & (J{+}1)B\\
-JB &  JA &  0 &  0 &  0 &  0 & -JB & -JA\\
 0 &  0 &  -(J{+}1)A &  -JB &  -(J{+}1)A &  -JB &  0 &  0\\
 0 &  0 &  -(J{+}1)B & JA & (J{+}1)B &  -JA &  0 &  0
\end{array}$}
\right],\lab{R8}\\
\s S &=& \left[\mbox{\scriptsize$
\begin{array}{cccccccc}
 0 &  0 &  0 &  0 &  0 &  0 & 2 &   0 \\
 0 &  0 &  0 &  0 &  0 &  0 &  0 & -1 \\
 0 &  0 &  0 &  0 &-B^2&-AB&  0 &  0\\
 0 &  0 &  0 &  0 &-AB  & -A^2&  0 &  0\\
 0 &  0 & -B^2  & -AB  &  0 &  0 &  0 &  0\\
 0 &  0 &  -AB & -A^2 &  0 &  0 &  0 &  0\\
2 &  0 &  0 &  0 &  0 &  0 &  0 &  0\\
 0 &  -1 &  0 &  0 &  0 &  0 &  0 &  0
\end{array}$}
\right]\lab{R9}
\end{eqnarray}
for $P=(-)^{J\pm1}$, and
%
\begin{eqnarray}
\s H &=& \left[\mbox{\scriptsize$
\begin{array}{cccccccc}
 0 &  0 &  -A & B &  -A &  -B &  0 &  0\\
 0 &  0 & B & A & B &  -A &  0 &  0\\
 A & -B &  0 &  0 &  0 &  0 &  A & -B\\
-B & -A &  0 &  0 &  0 &  0 &  B &  A\\
 A & -B &  0 &  0 &  0 &  0 &  A & -B\\
 B &  A &  0 &  0 &  0 &  0 & -B & -A\\
 0 &  0 & -A & -B & -A & B &  0 &  0\\
 0 &  0 & B &  -A &  B  & A &  0 &  0
\end{array}$}
\right],\lab{R10}\\
\s G &=& \left[\mbox{\scriptsize$
\begin{array}{cccccccc}
 0 &  0 & (J{+}1)A &  -(J{+}1)B & (J{+}1)A & (J{+}1)B &  0 &  0\\
 0 &  0 & JB & JA & JB &  -JA &  0 &  0\\
 (J{+}1)A &  JB &  0 &  0 &  0 &  0 &  (J{+}1)A &  JB\\
-(J{+}1)B &  JA &  0 &  0 &  0 &  0 &  (J{+}1)B & -JA\\
 (J{+}1)A &  JB &  0 &  0 &  0 &  0 &  (J{+}1)A &  JB\\
 (J{+}1)B & -JA &  0 &  0 &  0 &  0 & -(J{+}1)B &  JA\\
 0 &  0 &  (J{+}1)A & (J{+}1)B & (J{+}1)A &  -(J{+}1)B &  0 &  0\\
 0 &  0 & JB &  -JA & JB & JA &  0 &  0
\end{array}$}
\right],\lab{R11}\\
\s S &=& \left[\mbox{\scriptsize$
\begin{array}{cccccccc}
 0 &  0 &  0 &  0 &  0 &  0 & -B^2& -AB\\
 0 &  0 &  0 &  0 &  0 &  0 &  -AB & -A^2\\
 0 &  0 &  0 &  0 &  2 &  0 &  0 &  0\\
 0 &  0 &  0 &  0 &  0 & -1 &  0 &  0\\
 0 &  0 &  2 &  0 &  0 &  0 &  0 &  0\\
 0 &  0 &  0 & -1 &  0 &  0 &  0 &  0\\
-B^2& -AB &  0 &  0 &  0 &  0 &  0 &  0\\
-AB & -A^2&  0 &  0 &  0 &  0 &  0 &  0
\end{array}$}
\right]\lab{R12}
\end{eqnarray}
for $P=(-)^{J}$, where $A$, $B$ and $C$ are defined in Eq. \re{6.14}. Due
to the properties
%
\begin{equation}\lab{R14}
\s H^T = - \s H , \qquad  \s U^T = \s U ,
\end{equation}
the ``radial'' Hamiltonian ${\cal H}$ is a Hermitian operator with respect
to the inner product
%
\begin{equation}\lab{R15}
\langle\s Y|\s X\rangle_8 = \int\limits_0^\infty dr \s Y^\dagger(r)\s X(r) ,
\end{equation}
where the subscript ``8" denotes the dimensions of the vector-functions
$\s X,\s Y$.

	In the subsequent reduction of the set \re{R4} one can use the fact
that ${\rm rank}\,(\s H) = 4$ (for either parity). Thus one can reduce
the number of differential equations from 8 to 4. We perform this
reduction in a way that ensures, as far
as possible, the Hamiltonian structure of the equations.

	First of all we perform the orthogonal transformation:
%
\begin{equation}\lab{R16}
\tilde{\s X}(r) = \s E\s X(r), \qquad
\tilde{\cal H} = \s E{\cal H}\s E^{-1},
\end{equation}
where
%
\begin{eqnarray}
\s E &=& \frac{1}{\sqrt{2}}\left[\mbox{\scriptsize$
\begin{array}{cccccccc}
 1 &  0 &  0&  0 &  0 &  0 &  1 &  0\\
 0 & 0 &  A &  -B &  A & -B &  0 &  0\\
 0 & -1 &  0 &  0 &  0 &  0 &  0 &  1\\
 0 &  0 &  B &  A & -B & -A &  0 &  0\\
-1 &  0 &  0 &  0 &  0 &  0 &  1 &  0\\
 0 &  1 &  0 &  0 &  0 &  0 &  0 &  1\\
 0 &  0 &  B &  A &  B &  A &  0 &  0\\
 0 &  0 &  A &  -B & -A &  B &  0 &  0
\end{array}$}
\right],\quad P=(-)^{J\pm1},\lab{R17}\\
\s E &=& \frac{1}{\sqrt{2}}\left[\mbox{\scriptsize$
\begin{array}{cccccccc}
 0 &  0 &  1 &  0 &  1 &  0 & 0 & 0\\
 -A & B &  0 &  0 &  0 &  0 & -A & B\\
 0 &  0 &  0 & -1 &  0 &  1 &  0 &  0\\
-B & -A & 0 &  0 & 0 &  0 &  B & A\\
 B &  A &  0 &  0 &  0 &  0 &  B &  A\\
 A & -B &  0 &  0 &  0 &  0 & -A &   B\\
 0 &  0 & -1 &  0 &  1 &  0 &  0 &  0\\
 0 &  0 &  0 &  1 &  0 &  1 &  0 &  0
\end{array}$}
\right],\quad P=(-)^{J}\lab{R18}
\end{eqnarray}
It preserves the inner product \re{R15} and reduces the equations \re{R4}
to the form:
%
\begin{equation}\lab{R19}
\tilde{\cal H}\tilde{\s X}(r) \equiv \left\{\tilde{\s H} \frac{d}{dr} +
\tilde{\s U}(r)\right\}\tilde{\s X}(r) = E\tilde{\s X}(r),
\qquad \tilde{\s U} = {\s E} {\s U} {\s E}^{-1}.
\end{equation}

	It is convenient at this stage to express the 8-dimensional
vectors and matrices in terms of 4-dimensional blocks:
%
\begin{equation}\lab{R20}
\tilde{\s X} =
\left[
\begin{array}{c}
\tilde{\s X}_1\\
\tilde{\s X}_2
\end{array}
\right], \qquad
\tilde{\s V}(r) \equiv \tilde{\s U}(r) - E{\s I}=
\left[
\begin{array}{cc}
\tilde{\s V}_{11}(r)&\tilde{\s V}_{12}(r)\\
\tilde{\s V}_{21}(r)&\tilde{\s V}_{22}(r)
\end{array}
\right], \qquad {\it etc.}
\end{equation}
Then the matrix $\tilde{\s H}$ takes the form
%
\begin{equation}\lab{R21}
\tilde{\s H} = \left[
\begin{array}{cc}
\tilde{\s H}_{11} &\s 0\\ \s 0 &\s 0
\end{array}\right],
\qquad
\tilde{\s H}_{11} = 2\left[
\begin{array}{cc}
\s J &\s 0\\ \s 0 &\s J
\end{array}\right],
\qquad
\s J = \left[
\begin{array}{cc}
0 & -1\\ 1 &0
\end{array}\right],
\end{equation}
and the set of Eqs. \re{R19} becomes
%
\begin{eqnarray}
\tilde{\s H}_{11}\tilde{\s X}'_1(r) +
\tilde{\s V}_{11}(r)\tilde{\s X}_1(r) +
\tilde{\s V}_{12}(r)\tilde{\s X}_2(r) &=& 0,
\lab{R22}\\
\tilde{\s V}_{21}(r)\tilde{\s X}_1(r) +
\tilde{\s V}_{22}(r)\tilde{\s X}_2(r) &=& 0.
\lab{R23}
\end{eqnarray}

	The set \re{R23} is purely algebraic. It permits us to express
$\tilde{\s X}_2(r)$ in the terms of $\tilde{\s X}_1(r)$:
%
\begin{equation}\lab{R24}
\tilde{\s X}_2(r) =
-\tilde{\s V}_{22}^{-1}(r)\tilde{\s V}_{21}(r)\tilde{\s X}_1(r).
\end{equation}
Substitution of \re{R24} into \re{R22} yields the closed set of four
first-order differential equations:
%
\begin{equation}\lab{R25}
{\cal L}(E)\tilde{\s X}_1(r) \equiv
\left\{\frac12\tilde{\s H}_{11}\frac{d}{dr} + \tilde{\s
W}(E)\right\}\tilde{\s X}_1(r)  = 0,
\end{equation}
where
%
\begin{eqnarray}\lab{R26}
\tilde{\s W}(E)&\equiv&
(\tilde{\s V}_{11} -
\tilde{\s V}_{12}\tilde{\s V}_{22}^{-1}\tilde{\s V}_{21} )/2
\nonumber\\
&=&\frac12\left[
\begin{array}{cccc}
\begin{array}{lcr}
\lefteqn{\scriptstyle{-(E+3\alpha/r)}}&&\\
&\lefteqn{\mbox{\large$\scriptstyle{+\frac{m_\pm^2}{E-\alpha/r}}$}}&
\vspace{0.2em}\\
&&+\frac{4C^2/r^2}{E}
\end{array}
&
-2/r
&
0
&
-\frac{2Cm_\mp/r}{E}
\\&&&\\
-2/r
&
\begin{array}{lr}
\lefteqn{\scriptstyle{-(E+\alpha/r)}}&\\
&+\frac{m_\mp^2}{E+\alpha/r}
\end{array}
&
-\frac{2Cm_\mp /r}{E+\alpha/r}
&
0
\\&&&\\
0
&
-\frac{2Cm_\mp/r}{E+\alpha/r}
&
\begin{array}{lcr}
\mbox{\scriptsize$\lefteqn{-(E+2\alpha/r)}$}&&\\
&\lefteqn{\mbox{\large$\scriptstyle{+\frac{m_\pm^2}{E}}$}}&
\vspace{0.2em}\\
&&+\frac{4C^2/r^2}{E+\alpha/r}
\end{array}
&
0
\\&&&\\
-\frac{2Cm_\mp/r}{E}
&
0
&
0
&
\begin{array}{lr}
\mbox{\scriptsize$\lefteqn{-(E+2\alpha/r)}$}&\\
&+\frac{m_\mp^2}{E}
\end{array}
\end{array}
\; \; \right].
\end{eqnarray}
Here the upper sign corresponds to $P=(-)^{J+1}$, and the lower sign
corresponds to $P=(-)^{J}$.

The operator  ${\cal L}(E)$ in Eq. \re{R25} is
formally Hermitian, {\sl i.e.}, given $E$, it is Hermitian with respect to
the inner product $\langle\dots|\dots\rangle_4$. But any two solutions of
\re{R25}, ${\s X}_1$ and ${\s Y}_1$,  corresponding to different values of
the energy, $E$ and $E^\prime$, are not orthogonal. This is due to the
nonlinear dependence of ${\cal L}(E)$ on $E$. Orthogonality can be instated
by using the following definition of inner product:
%
\begin{equation}\lab{R26a}
\langle\langle {\s Y}_1|{\s X}_1\rangle\rangle_4 =
\left\langle {\s Y}_1\left|
\frac{{\cal L}(E^\prime) - {\cal L}(E)}{E^\prime - E}
\right|{\s X}_1\right\rangle_4.
\end{equation}
This inner product follows directly by
substitution of Eq. \re{R24} into Eq. \re{R15}.

	The set of first order equations \re{R25} can also be expressed as a
second order equation. For this purposeit is convenient to permute the
elements of ${\s X}_1$ by means of the matrix ${\s L}$, where
%
\begin{equation}
\s L = \left[\mbox{\scriptsize$
\begin{array}{cccc}
 1 &  0 &  0 &  0 \\
 0 &  0 &  1 &  0 \\
 0 &  1 &  0 &  0 \\
 0 &  0 &  0 &  1
\end{array}$}
\right] \quad {\rm for} \quad P=(-)^{J\pm1},\lab{R27}
\end{equation}
and
\begin{equation}
\s L = \left[\mbox{\scriptsize$
\begin{array}{cccccccc}
 0 &  1 &  0 &  0 \\
 0 &  0 &  0 & -1 \\
-1 &  0 &  0 &  0 \\
 0 &  0 &  1 &  0
\end{array}$}
\right] \quad {\rm for} \quad P=(-)^{J} . \lab{R28}
\end{equation}
Then, in  terms of the two-dimensional blocks
%
\begin{equation}\lab{R29}
\bar{\s X}_1  = \s L\tilde{\s X}_1
\equiv \left[
\begin{array}{c}
\Psi_1\\
\Psi_2
\end{array}
\right] ,
\end{equation}
the equations take the form
%
\begin{eqnarray}
-\Psi_2^\prime + \bar{\s W}_{11}\Psi_1+
\bar{\s W}_{12}\Psi_2 &=& 0,
\lab{R30}\\
\Psi_1^\prime + \bar{\s W}_{21}\Psi_1+
\bar{\s W}_{22}\Psi_2 &=& 0,
\lab{R31}
\end{eqnarray}
where $\bar{\s W} = {\s L} \, {\tilde {\s W}} \, {\s L}^{-1}$.
Elimination of $\Psi_2$ leads to the following equation for the $2 \times
1$  vector-function $\Psi_1$:
%
\begin{equation}\lab{R32}
{\cal F}(E) \equiv
\left[\left\{\frac{d}{dr} - \bar{\s W}_{12}\right\}\bar{\s W}_{22}^{-1}
\left\{\frac{d}{dr} + \bar{\s W}_{21}\right\} + \bar{\s
W}_{11}\right]\Psi_1(r) = 0.
\end{equation}
Eigenstates $\Psi_1$, $\Phi_1$ corresponding to different values of the
energy are orthogonal with respect to the inner product:
%
\begin{equation}\lab{R32a}
\langle\langle \Phi_1|\Psi_1\rangle\rangle_2 =
\left\langle \Phi_1\left|
\frac{{\cal F}(E^\prime) - {\cal F}(E)}{E^\prime - E}
\right|\Psi_1\right\rangle_2 ,
\end{equation}
which also follows from the reduction procedure.


\setcounter{equation}{0}
\renewcommand{\theequation}{11-\arabic{equation}}

\section{Perturbative solutions for $J>0$ states}

The form of equations \re{R32} is convenient for examining the  energy
spectrum of $J>0$ bound states perturbatively to $O(\alpha^4)$. For  this
purpose we introduce the dimensionless quantities,
%
\begin{equation}\lab{P1}
\rho = \mu\alpha r,\qquad \lambda = \frac{E-m_+}{\mu\alpha^2}, \qquad
{\rm and} \qquad \delta = \frac{m_-}{m_+},
\end{equation}
where $\mu=m_1m_2/m_+$ is the reduced mass. We now perform a perturbative
expansion in $\alpha$ of equation \re{R32}.  To order $\alpha^2$
equation \re{R32} takes the Hamiltonian form (hereafter we omit the
subscript ``1" of the wave function $\Psi_1$):
%
\begin{equation}\lab{P2}
{\cal H}\Psi(\rho) \approx \left\{{\cal H}^{(0)} +
\alpha^2{\cal H}^{(1)}(\lambda)\right\}\Psi(\rho)=
\lambda\Psi(\rho),
\end{equation}
where
%
\begin{equation}\lab{P3}
\Psi(\rho) =  \left[\begin{array}{c}
\psi_1(\rho)\\
\psi_2(\rho)
\end{array}
\right]
\end{equation}
is a two-component wave function, ${\cal H}$ is the Hamiltonian divided by
$\mu\alpha^2$ and expressed in terms of dimensionless quantities \re{P1},
%
\begin{equation}\lab{P4}
{\cal H}^{(0)} = -\frac12\left\{\frac{d}{d\rho^2} -  \frac{1}{\rho^2}{\cal
J}\right\} - \frac{1}{\rho}
\end{equation}
is the unperturbed (i.e., $0$th-order) Hamiltonian, and
%
\begin{equation}
{\cal H}^{(1)}(\lambda) =  \frac{d}{d\rho}{\cal
K}(\rho,\lambda)\frac{d}{d\rho} +{\cal M}(\rho,\lambda)
\lab{P5}
\end{equation}
is the perturbative correction to \re{P4}. The form of the symmetric
$2\times2$ matrices ${\cal J}$, ${\cal K}(\rho,\lambda)$, ${\cal
M}(\rho,\lambda)$ depends on the parity. In order to obtain the energy
spectrum to $O(\alpha^4)$, it is sufficient to calculate the eigenvalues of
$\lambda$ to $O(\alpha^2)$, {\sl i.e.}, $\lambda \approx \lambda^{(0)} +
\alpha^2\lambda^{(1)}$, where $\lambda^{(0)}$ will be calculated exactly,
while for $\lambda^{(1)}$ first order perturbation theory in $\alpha^2$
is sufficient. Hence, the dependence of ${\cal H}^{(1)}(\lambda)$ on
$\lambda$ is not crucial: to the accuracy required, it can be replaced
by$\lambda^{(0)}$. In addition, the kernel of the inner product \re{R32a}
can  be set to unity.

In the case $P=(-)^{J\pm1}$ we have ${\cal J}= C^2\s I$,  so that  ${\cal
H}^{(0)}$ is the (dimensionless) radial Coulomb Hamiltonian $H_J$  with the
angular momentum $\ell=J$, repeated twice:
%
\begin{equation}\lab{P6}
{\cal H}^{(0)}=
\left[\mbox{\scriptsize$
\begin{array}{cc}
H_{J} & 0\\
0 &H_{J}
\end{array}$}
\right], \qquad H_J = -{1 \over 2} \left\{\frac{d}{d\rho^2} -
\frac{J(J+1)}{\rho^2}\right\} - \frac{1}{\rho} .
\end{equation}
The matrices ${\cal K}(\rho,\lambda)$ and ${\cal M}(\rho,\lambda)$ are
%
\begin{eqnarray}
{\cal K}(\rho,\lambda)&=&
\frac18\left[\mbox{\scriptsize$
\begin{array}{cc}
(1+\delta^2)(\lambda+\frac1{\rho}) & 0\\
0 & (1+\delta^2)\lambda+\frac2{\rho}
\end{array}$}
\right],\lab{P7}\\
{\cal M}(\rho,\lambda)&=&
\frac18\left[\mbox{\scriptsize$
\begin{array}{cc}
(1-\delta^2)\lambda(\lambda-\frac2{\rho}) +\frac{1-\delta^2 - C^2\lambda
(1+\delta^2)}{\rho^2}+\frac{1+\delta^2(1 -2C^2)}{\rho^3}&
\frac{2C\delta}{\rho^3}\\
\frac{2C\delta}{\rho^3}&
(1-\delta^2)\lambda^2 - \frac{C^2(1+\delta^2)}{\rho^2}(\lambda+\frac1{\rho})
\end{array}$} \! \right] \lab{P8}
\end{eqnarray}

	The eigenvalues of the $0$th-order Hamiltonian \re{P6}, namely
%
\begin{equation}\lab{P9}
\lambda^{(0)}=-1/(2n^2),\qquad n=1,2,\dots,
\end{equation}
are two-fold degenerate, each with the two eigenstates
%
\begin{equation}\lab{P10}
\Psi_{(1)}^{(0)} = \left[\begin{array}{c}
|n,J\rangle\\
0
\end{array}
\right]
\qquad {\rm and} \qquad
\Psi_{(2)}^{(0)} = \left[\begin{array}{c}
0\\
|n,J\rangle
\end{array}
\right],
\end{equation}
where $|n,J\rangle$ is a solution of the Coulomb problem $H_{J}|n,J\rangle
= \lambda^{(0)}|n,J\rangle$. Thus, the correction $\lambda^{(1)}$ must be
calculated appropriately for the degenerate situation:
%
\begin{equation}\lab{P11}
\lambda^{(1)}_{(1,2)} = \frac12\left[\Lambda_{11} + \Lambda_{22} \pm
\sqrt{(\Lambda_{11} - \Lambda_{22})^2 + 4 \Lambda_{12}^2}\right],
\end{equation}
where the matrix $\Lambda$ is defined as follows:
%
\begin{eqnarray}
\Lambda &=& \left[ \langle \Psi_{(i)}^{(0)}| {\cal H}^{(1)}(\lambda^{(0)})
|\Psi_{(j)}^{(0)}\rangle \right] \nonumber\\ &=&
\left[\mbox{\scriptsize$
\begin{array}{cc}
\frac{11+\delta^2}{32n^4} -\frac{\delta^2}{(2J+1)n^3}&
\frac{\delta}{2C(2J+1)n^3}\\
\frac{\delta}{2C(2J+1)n^3}&
\frac{11+\delta^2}{32n^4} -\frac{(3+\delta^2)C^2 +2}{4C^2(2J+1)n^3}
\end{array}$}
\right].\lab{P12}
\end{eqnarray}

	The mass spectrum
%
\begin{equation}\lab{P13}
E_{(1,2)} = m_+ + \mu\alpha^2\lambda^{(0)} +
\mu\alpha^2\lambda^{(1)}_{(1,2)},
\end{equation}
obtained with the use of \re{P9}, \re{P11} and \re{P12} coincides neither
with the muonium spectrum found in \cite{Connell,Hersb}  nor (if $m_1=m_2$)
with the spectrum of parapositronium (see \cite{BS57} and refs. therein).
The reason, like for the $J=0$ states, lies in the use of the empty vacuum
\re{3.4}, which leads to  single-particle states of  positive and negative
energy, and subsequently to the Breit equation with  its spurious term
$H^\prime$ \re{8-7} in the Hamiltonian  \cite{Chrap53,Bar55,BS57}, as
discussed in connection with the $J=0$  states. We are going to show that
the elimination of the  contribution of $H^\prime$ from the spectrum leads
to the correct  result.

First of all we transform $H^\prime$ into a radial representation.  For
this purpose we note that $\psi_1=s_1+v_1$ ($\approx s_1$ in the
nonrelativistic limit), i.e., $\psi_1$ contains only those components of
$F(\B r)$ \re{6-1} which are coefficients of ``bispinor harmonics"
$\phi^A(\bar r)$. Similarly, $\psi_2=-s_2+v_2$ ($\approx -s_2$ in the
nonrelativistic limit), i.e., $\psi_1$ contains only coefficients of the
``bispinor harmonics" $\phi^0(\bar r)$. Thus, for $P=(-)^{J\pm1}$  parity
states the spurious term (divided by $\mu\alpha^4$) takes the  following
radial form:
%
\begin{equation}\lab{P14}
{\cal H}^\prime=\frac{1}{\mu\alpha^4}\left[ \int d\hat r \; {\rm
Tr}(\varphi_{i}^{\, \dag} \,H^\prime \varphi_{j})
\right]=\frac{1-\delta^2}{8\rho^2}\left[\mbox{\scriptsize$
\begin{array}{cc}
4 & 0\\
0 &1
\end{array}$}
\right] ,
\end{equation}
where $i,j=A,0$, and the corresponding matrix elements are
%
\begin{eqnarray}
\Lambda^\prime &=& \left[\langle\Psi_{(i)}^{(0)}| {\cal H}^\prime
|\Psi_{(j)}^{(0)}\rangle  \right] = \frac{1-\delta^2}{4(2J+1)n^3}
\left[\mbox{\scriptsize$
\begin{array}{cc}
4 & 0\\
0 &1
\end{array}$}
\right].
\lab{P15}
\end{eqnarray}
If we now  use $\bar\Lambda = \Lambda - \Lambda^\prime$, instead of
$\Lambda$, in \re{P11} and \re{P13}, we obtain the spectrum,
%
\begin{equation}\lab{P16}
E_{(1,2)} = m_+ - \frac{\mu\alpha^2}{2n^2} + \frac{\mu\alpha^4}{4n^3}
\left\{\frac{11+\delta^2}{8n} -
\frac{2J+1}{C^2} \pm \frac{\sqrt{1+4C^2\delta^2}}{(2J+1)C^2}\right\}
\end{equation}
which coincides with the results of Connell \cite{Connell} and Hersbach
\cite{Hersb} for the parity $(-1)^{J \pm 1}$ states. Thus, correcting
for the spurious terms in the Breit Hamiltonian, we obtain the expected
$O(\alpha^4)$ results.

	In the  $P=(-1)^{J}$ case, the matrix ${\cal J}$ is not diagonal:
%
\begin{equation}\lab{P17}
{\cal J}=
\left[\mbox{\scriptsize$
\begin{array}{cc}
C^2 +2 & -2C\\
-2C & C^2
\end{array}$}
\right].
\end{equation}
It can be diagonalized by means of the orthogonal transformation,  using
the matrix:
%
\begin{equation}\lab{P18}
{\cal R}=
\left[\mbox{\scriptsize$
\begin{array}{cc}
A & -B\\
B & A
\end{array}$}
\right].
\end{equation}
Then
%
\begin{equation}\lab{P19}
\tilde{\cal J}={\cal R}{\cal J}{\cal R}^{-1} =
\left[\mbox{\scriptsize$
\begin{array}{cc}
(J+1)(J+2) & 0\\
0 & (J-1)J
\end{array}$}
\right],
\end{equation}
so that the $0$th-order Hamiltonian becomes:
%
\begin{equation}\lab{P20}
\tilde{\cal H}^{(0)}={\cal R}{\cal H}^{(0)}{\cal R}^{-1} =
\left[\mbox{\scriptsize$
\begin{array}{cc}
H_{J+1} & 0\\
0 &H_{J-1}
\end{array}$}
\right].
\end{equation}
It possesses the doubly degenerate eigenvalues \re{P9} with eigenstates:
%
\begin{equation}\lab{P21}
\Psi_{(1)}^{(0)} = \left[\begin{array}{c}
|n,J+1\rangle\\
0
\end{array}
\right]
\qquad {\rm and} \qquad
\Psi_{(2)}^{(0)} = \left[\begin{array}{c}
0\\
|n,J-1\rangle
\end{array}
\right],
\end{equation}

The first-order correction $\tilde{\cal H}^{(1)}$ \re{P5}, with the
matrices
%
\begin{equation}
\tilde{\cal K}(\rho,\lambda)=
\frac18\left[\mbox{\scriptsize$
\begin{array}{cc}
(1+\delta^2)\lambda + \frac{(3-\delta^2)A^2+2B^2}{\rho} -
\frac{C^2(1-\delta^2)}{\rho^2}&\frac{AB(1-\delta^2)}{\rho} \\
\frac{AB(1-\delta^2)}{\rho} &
(1+\delta^2)\lambda + \frac{2A^2+(3-\delta^2)B^2}{\rho} -
\frac{C^2(1-\delta^2)}{\rho^2}
\end{array}$}
\right],\lab{P22}
\end{equation}
\begin{equation}
\tilde{\cal M}(\rho,\lambda)=
\frac18\left[\mbox{\scriptsize$
\begin{array}{lr}
\begin{array}{llll}
\hspace{-1em}
\lefteqn{\scriptstyle{
(1-\delta^2)\lambda(\lambda + 2A^2/\rho)
}}&&&\\
&\hspace{-1em}
\lefteqn{\scriptstyle{
-A^2[\{\lambda(6J^2+11J+2)-1\} -  \delta^2\{\lambda(2J^2+J-2)-1\}]/\rho^2
}}&&\\
&&\hspace{-1em}
\lefteqn{\scriptstyle{
-A^2[7J^2+22J+9 - \delta^2(3J^2+8J+3)]/\rho^3
}}&\\
&&&\hspace{-1em}
\lefteqn{\scriptstyle{
+ (1-\delta^2)J(J+1)^2(J+4)/\rho^4
}}\hspace{13em}
\end{array}
&
\frac{AB(1-\delta^2)}{\rho}\left(2\lambda+\frac{1+2\lambda}{\rho} -
\frac{C^2}{\rho^2}\right)\\
\frac{AB(1-\delta^2)}{\rho}\left(2\lambda+\frac{1+2\lambda}{\rho} -
\frac{C^2}{\rho^2}\right)
&
\begin{array}{llll}
\lefteqn{\scriptstyle{
(1-\delta^2)\lambda(\lambda + 2B^2/\rho)
}}&&&\\
&
\lefteqn{\scriptstyle{
-B^2[\{\lambda(6J^2+J-3)-1\} -  \delta^2\{\lambda(2J^2+3J-1)-1\}]/\rho^2
}}&&\\
&&
\lefteqn{\scriptstyle{
-B^2[7J^2-8J-6 - \delta^2(3J^2-2J-2)]/\rho^3
}}&\\
&&&
\lefteqn{\scriptstyle{
+ (1-\delta^2)J(J+1)^2(J-3)/\rho^4
}}\hspace{16em}
\end{array}
\end{array}$}
\right] , \lab{P23}
\end{equation}
generates diagonal matrix elements only:
%
\begin{eqnarray}
\Lambda &=& \left[ \langle \Psi_{(i)}^{(0)}|  \tilde{\cal
H}^{(1)}(\lambda^{(0)})  |\Psi_{(j)}^{(0)}\rangle \right] \nonumber\\ &=&
\left[\mbox{\scriptsize$
\begin{array}{cc}
\frac{11+\delta^2}{32n^4} -
\frac{(7+\delta^2)J^2+(17-\delta^2)J+8-2\delta^2}{4(2J+1)(J+1)(2J+3)n^3}&
0\\
0&
\frac{11+\delta^2}{32n^4} -
\frac{(7+\delta^2)J^2-3(1-\delta^2)J-2}{4(2J-1)J(2J+1)n^3}
\end{array}$}
\right].\lab{P24}
\end{eqnarray}
The energy spectrum \re{P13} with $\lambda^{(1)}_{(1)}=\Lambda_{11}$,
$\lambda^{(1)}_{(2)}=\Lambda_{22}$  contains the contribution of the
spurious term \re{8-7}. Again, we present this term in radial form:
%
\begin{equation}\lab{P25}
{\cal H}^\prime=\frac{1}{\mu\alpha^4}\left[ \int d\hat r \; {\rm
Tr}(\varphi_{i}^{\, \dag} \,H^\prime \varphi_{j}) \right]
\nonumber\\
=\frac{1-\delta^2}{8\rho^2}
\left[\mbox{\scriptsize$
\begin{array}{cc}
B^2 & AB\\
AB & A^2
\end{array}$}
\right] ,
\end{equation}
where, for the present $P=(-)^{J\pm1}$ parity case, $i,j=-,+$. The
corresponding matrix elements are:
%
\begin{eqnarray}
\Lambda^\prime &=& \left[  \langle \Psi_{(i)}^{(0)}| {\cal H}^\prime
|\Psi_{(j)}^{(0)}\rangle  \right] = \frac{1-\delta^2}{4n^3}
\left[\mbox{\scriptsize$
\begin{array}{cc}
\frac{B^2}{2J+3} & 0\\
0 & \frac{A^2}{2J-1}
\end{array}$}
\right].
\lab{P26}
\end{eqnarray}
The substitution of $\lambda^{(0)}_{(1)}=\Lambda_{11} -
\Lambda^\prime_{11}$ and $\lambda^{(0)}_{(2)}=\Lambda_{22} -
\Lambda^\prime_{22}$ into \re{P13}  yields the spectrum:
%
\begin{equation}\lab{P27}
E_{(1,2)} = m_+ - \frac{\mu\alpha^2}{2n^2} +
\mu\alpha^4\frac{11+\delta^2}{32n^4} -
\frac{\mu\alpha^4}{2n^3}\times\left\{\displaystyle{
\frac{1}{J+1}+\frac{1-\delta^2}{(2J+3)(2J+1)}} \atop \displaystyle{
\frac{1}{J}-\frac{1-\delta^2}{(2J+1)(2J-1)}}\right.
\end{equation}
which coincides with the results of Connell \cite{Connell} and Hersbach
\cite{Hersb} for the parity $(-1)^{J}$ states.


\setcounter{equation}{0}
\renewcommand{\theequation}{12-\arabic{equation}}

\section{Concluding Remarks}

We have studied a reformulation of QED, in which the coupled
Dirac-Maxwell field equations are partially decoupled by expressing the
mediating photon field in terms of the Dirac particle field, using
covariant Green's functions. This allows us to reformulate the Hamiltonian
of the theory so that the photon propagator appears directly in a quartic,
nonlocal interaction term. We then consider a truncated model, in which
there are no free (physical) photons. For such a model, each  $N$-particle
segment of the Fock space of the quantized, equal-time Hamiltonian is an
invariant space,  that is, there is no coupling among the various
$N$-fermion segments. This is achieved by introducing an unconventional
``empty''   vacuum state. As a consequence, there exist exact few-particle
eigenstates of the truncated Hamiltonian, which lead to
 Dirac-like two- and three-fermion wave equations.
We show, in particular, that the  two-fermion wave equation, in
the Coulomb gauge, is just the Breit equation.

For specific $J^P$ states, the Breit equation reduces to  radial
form, and then to Dirac-like equations for $J=0$ states, and to a
coupled pair of Schr\"odinger-like equations for $J>0$ states.
Perturbative solution of these equations yields $\alpha^4$-corrections
to the nonrelativistic Rydberg spectrum, which do not reproduce
the muonium spectrum as calculated
by Connell \cite{Connell} and Hersbach \cite{Hersb}
  (nor the positronium spectrum in the $m_1=m_2$ case).
 The  apparent reason for this disagreement is the mixing of
positive- and negative-energy states, which is characteristic of
the Breit equation \cite{BR51,Chrap53,BS57}. However, agreement
is achieved if we subtract the contribution of spurious operator
\re{8-7}, which appears in the Breit equation ({\it c.f.}
Eqs. \re{8-9}, \re{8-5}, \re{P16},  \re{P27}).

We have not been able to obtain analytic solutions of the radial
equations.
	These radial wave equations have, in general, a
singular point at  $r_1\sim\alpha/E > 0$, where $E$ is the two-fermion
bound state energy (rest mass). The existence of such an
 ``interior'' singularity makes it difficult to obtain numerical
solutions of the radial boundary value problem by standard methods.
 The only exception  is  the case
$J=0^+$, $m_1=m_2$, for which the radial equations
are regular, and which we have therefore  studied numerically.
(It is noteworthy that this is the case
where the contribution of the operator \re{8-7} is zero.)

Our numerical results for the equal-mass  $J=0^+$ states, show that
the dependence of the energy $E$ on the coupling constant
$\alpha$ is  qualitatively similar to that obtained earlier
for the Coulomb QED model (for which transverse-photon interactions
are ignored)  \cite{DdiL96}. For low $\alpha$, the numerically
obtained eigenenergies are in agreement with the
result derived perturbatively.  Thereafter, $E(\alpha)$ decreases
monotonically to $E(\alpha_c)>0$ as  $\alpha$ approaches a critical
value $\alpha_c$. We find that $\alpha_c = 2/\sqrt{3}$ (in contrast to CQED
value of $\alpha_c=2$).
\vfill \eject


\end{document}